\documentclass[pdflatex,sn-mathphys-num]{sn-jnl}

\usepackage{graphicx}%
\usepackage{multirow}%
\usepackage{amsmath,amssymb,amsfonts}%
\usepackage{amsthm}%
\usepackage{mathrsfs}%
\usepackage[title]{appendix}%
\usepackage{xcolor}%
\usepackage{textcomp}%
\usepackage{manyfoot}%
\usepackage{booktabs}%
\usepackage{algorithm}%
\usepackage{algorithmicx}%
\usepackage{algpseudocode}%
\usepackage{listings}%

\theoremstyle{thmstyleone}%
\theoremstyle{thmstyletwo}%

\theoremstyle{thmstylethree}%

\begin{document}

\title[Article Title]{Multi-View Adaptive Contrastive Learning for Information Retrieval Based Fault Localization}


\author[1]{\fnm{Chunying} \sur{Zhou}}

\author*[1]{\fnm{Xiaoyuan} \sur{Xie}}\email{xxie@whu.edu.cn}

\author[1]{\fnm{Gong} \sur{Chen}}

\author[2]{\fnm{Peng} \sur{He}}

\author[1]{\fnm{Bing} \sur{Li}}


\affil[1]{\orgdiv{School of Computer Science}, \orgname{Wuhan University}, \orgaddress{\city{Wuhan}, \postcode{430072}, \country{China}}}
\affil[2]{\orgdiv{School of Cyber Science and Technology}, \orgname{Hubei University}, \orgaddress{\city{Wuhan}, \postcode{430062}, \country{China}}}


\abstract{Most studies focused on information retrieval-based techniques for fault localization, which built representations for bug reports and source code files and matched their semantic vectors through similarity measurement. However, such approaches often ignore some useful information that might help improve localization performance, such as 1) the interaction relationship between bug reports and source code files; 2) the similarity relationship between bug reports; and 3) the co-citation relationship between source code files. 
In this paper, we propose a novel approach named \textbf{M}ulti-View \textbf{A}daptive \textbf{C}ontrastive \textbf{L}earning for \textbf{I}nformation \textbf{R}etrieval \textbf{F}ault \textbf{L}ocalization (MACL-IRFL) to learn the above-mentioned relationships for software fault localization. 
Specifically, we first generate data augmentations from report-code interaction view, report-report similarity view and code-code co-citation view separately, and adopt graph neural network to aggregate the information of bug reports or source code files from the three views in the embedding process. Moreover, we perform contrastive learning across these views. Our design of contrastive learning task will force the bug report representations to encode information shared by report-report and report-code views, and the source code file representations shared by code-code and report-code views, thereby alleviating the noise from auxiliary information.
Finally, to evaluate the performance of our approach, we conduct extensive experiments on five open-source Java projects. The results show that our model can improve over the best baseline up to 28.93\%, 25.57\% and 20.35\% on Accuracy@1, MAP and MRR, respectively.}




\maketitle

\section{Introduction}\label{sec1}

A project team could receive a large number of bug reports during the life cycle of a software system. Once a bug report is received, a developer needs to be assigned to review source code files and locate the buggy files to fix them. However, the process of developers analyzing bug reports and a lot of source code files is undoubtedly labor-intensive and time-consuming. To reduce the cost of software maintenance and improve the efficiency of software debugging, it is desirable to use fault localization techniques to help software engineers automatically locate faults.

Existing approaches can be categorized into two mainstreams: 1) Spectrum-based approaches \cite{yoo2017human, tu2019analysis, natan2023distributed, yan2023fault}, which require dynamic spectra information between different executions of the program to help locate faults. However, it is hard to collect many passing and failing execution traces in software maintenance phase. 2) Information retrieval (IR)-based approaches \cite{zhou2012should, ye2014learning}, which do not require dynamic execution information, but analyze the text of bug reports and source code files to measure the relevance between them. 
However, there is a significant lexical gap between the natural language in bug reports and the programming language in source code files \cite{ye2014learning}. To bridge the lexical gap, researchers designed a series of semantic matching models utilizing deep neural network and IR techniques, and improved localization performance \cite{ye2016word, lam2015combining, lam2017bug, gharibi2018leveraging, xiao2017improving, xiao2019improving, qi2021dreamloc, zhang2020exploiting, liang2022modeling, han2023bjxnet}.

Although semantic matching based models have made some progress, the performance of IR-based approaches is often affected by the textual quality of bug reports.
Specifically, it is very difficult to have satisfactory performance when insufficient textual description is provided in bug reports, even with very sophisticated models.
In fact, apart from mining the relationship within an individual pair of bug report and code file, there are some other auxiliary information that can facilitate IR-based fault localization.
For example, 1) the similarity relationships between bug reports. It is possible to receive duplicate or similar bug reports in a defect tracking system \cite{chaparro2017improving}. 
Previous works \cite{zhou2012should,davies2012using,li2021survey} show that if two bug reports are similar, they are likely to be resulted by similar buggy files. 
Thus, we can adopt such similarity relationships to recommend potential candidates. And 2) the co-citation relationships between code files. In many cases, multiple files are co-cited in one bug report, and are coupled in the corresponding bug fixing. Such a coupling relation may be intrinsically due to the software code dependency (for example, calling hierarchy), therefore may also exist in other bug fixing and the coupled files are located together consequently.

\begin{figure}
	\centering
	  \includegraphics[width=0.7\linewidth]{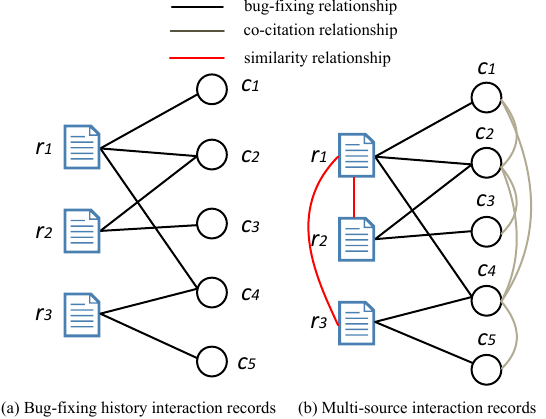}
	\caption{Illustrate the multi-relationships between bug reports and source code files, where $r_i$ denotes the bug reports, $c_i$ denotes the code files.}
	\label{example}
\end{figure}

These auxiliary information can be leveraged to better locate in the case of insufficient bug textual information \cite{li2021survey}. However, existing studies have not deeply explored the impact of multiple implicit information on location performance, and how to optimize the adaptive extraction of valuable information from multiple sources.
All the above discussion motivates us to design a new automatic fault localization approach, which utilizes auxiliary information to supplement the insufficient bug textual information, and automatically encodes different information into the representation learnings of bug reports and source code files. 

Fig. \ref{example}(a) shows the report-code bug-fixing history interaction relationships modeled as a typical bipartite graph structure. In practice, there are lots of implicit relationships among bug reports and source code files, which might help better locate the bug-relevant files based on a given bug report.
We integrate the implicit relationships into the bipartite graph as shown in Fig. \ref{example}(b). 
For example, in Fig. \ref{example}(b), as bug report $r_{1}$, the relevant buggy files are $c_{1}$,$c_{2}$,$c_{4}$, so there are explicit interactions between them (the black line). At the same time, there are co-citation relationships between $c_{1}$,$c_{2}$ and $c_{4}$ (the gray line). 
In addition, there exists a certain degree of similarity between $r_{1}$ and $r_{2}$ (or $r_{1}$ and $r_{3}$), as they share commonly fixed buggy files (the red line). To a certain extent, these relationships can help the localization process.

However, we argue that the core challenge of multiple sources fault localization are mainly reflected in two aspects.
\textit{Absence of bug-fixing history information during the prediction stage:} During model training, bug-fixing history records often play a pivotal role. 
Nevertheless, the validation and test sets comprise new bug reports that inherently lack these fixing history records, resulting in a disconnection between new bug reports and the source code files during the validation and test phases. 
Hence, we must explore supplementary information sources to forge connections between the new reports and the nodes encountered during the training phase. \textit{Auxiliary information overload:} The auxiliary information such as the similarity relationships between bug reports and the co-citation relationships between code files contains many irrelevant information. The risk lies in the model will inevitably enlarge the impact of such noises, potentially overshadowing the effect of useful information and compromising effective fault localization. 

Inspired by the recent success of contrastive learning (CL) technique, in this work, we propose Multi-View Adaptive Contrastive Learning (MACL-IRFL) fault localization method to address the above limitations. 
To tackle the \textit{Absence of bug-fixing history information during the prediction stage}, we model bug reports, source code files, and the relationships between them as structured graphs from three views (i.e., report-code interaction view, report-report similarity view, and code-code co-citation view). These views provide auxiliary information that supplements the missing historical connections in the validation and test stages. By performing message aggregation within each view, our approach effectively captures meaningful interactions across different types of data.
For the \textit{Auxiliary information overload} challenge, we leverage the power of contrastive learning. Specifically, the contrastive loss aligns the representations of a report across report-report view and report-code view (or a code file across code-code view and report-code view) to be closer to each other, while those of different reports (or code files) apart. This alignment enables the representations of bug reports and source code files to focus on the shared and relevant information between views while filtering out noise from irrelevant auxiliary data.

In a nutshell, our contributions summarize as:

(1) We formalize three types of relationships between bug reports and source code files, which include: 1) the bug-fixing history interaction relationship between bug reports and source code files; 2) the similarity relationship between bug reports; and 3) the co-citation relationship between source code files. We construct three views with the above three types of relationships and the two kinds of nodes, i.e., report-code view, report-report view, and code-code view. 

(2) We propose a novel model named MACL-IRFL, which takes advantage of contrastive learning to automatically drop task-irrelevant information and encode the useful information shared between report-code, report-report, and code-code views for highly node representations.

(3) To evaluate the performance of our approach, we conduct extensive experiments on five open-source projects. Experimental results show that our proposed approach outperforms other state-of-the-art baselines. 

The remainder of this paper is structured as follows. Section \ref{background} presents the background knowledge in this paper. The definitions and preliminary knowledge related to our framework are reviewed in Section \ref{definitions}. Section \ref{proposed framework} gives the overall framework and the detailed description of our approach. Section \ref{experiments} presents
experiment settings and evaluation metrics, and discusses the experimental results through three research questions. The threats to validity are discussed in Section \ref{threats to validity}. Section \ref{related work} presents the related works. We conclude this paper in Section \ref{conclusion}.

\section{Background}\label{background}
This work is mainly relevant to three research lines: IR-based fault localization, graph neural networks, and contrastive learning.

\subsection{IR-based Fault Localization}

IR-based approaches treat a bug report as a query and a source file as a document, and provide a recommendation list of candidate source files based on the textual similarity between the report and the source file. 
Early IR-based studies rely on the textual similarity between the bug reports and source code files to identify suspicious files using classical IR techniques such as Latent Dirichlet Allocation (LDA) \cite{lukins2008source, lukins2010bug, nguyen2011topic}, Vector Space Model (VSM) \cite{zhou2012should, saha2013improving}, Latent Semantic Analysis (LSA) \cite{cleary2009empirical}. The basic intuition for adopting these text-based approaches for fault localization is that bug reports and project resources share common words, which may falter in their effectiveness when the source code contains limited common terms with a new bug report, potentially hindering their ability to accurately locate faults.
Although there may not be substantial overlap in the shared common words, there is a high likelihood that they convey the same meaning semantically.  Consequently, to bridge the gap between natural language and programming code, extensive research \cite{ye2016word, lam2015combining, lam2017bug, ibrahim2023explainable, van2020review, gharibi2018leveraging, xiao2017improving, church2017word2vec, xiao2019improving} has been devoted to leveraging deep learning techniques for semantic matching.
Although these semantic-based approaches achieved promising performance, they directly used representation-focused models from natural language processing (NLP) tasks to learn the representation of bug reports and source files without considering their characteristics. To this end, \cite{huo2020control, zhang2020exploiting, han2023bjxnet} constructed source code as control flow graph (CFG), code knowledge graph, and code property graph (CPG) respectively, aiming to extract intricate structural information to  enrich the code characteristics. 
Furthermore, in addition to semantic matching, several studies have integrated other techniques to enhance the performance of fault localization. For instance, Zhu et al. \cite{zhu2022bl} employed a generative adversarial network (GAN) \cite{liu2019generative, creswell2018generative} model to learn from historical bug-fixing records and simulate the generation of new fix proposals for localization.  

For short, the primary goal of these IR-based approaches is to learn more accurate and rich representations of bug reports and source code to facilitate better matching. However, the performance of these approaches can be impacted by the limited content and quality of bug reports. To mitigate this issue, we explore auxiliary information (report-report similarity relationships and code-code co-citation relationships) and incorporate historical bug-fixing records of bug reports to construct three views, and adopt contrastive learning strategy to learn interactive representations between bug reports and source code from the three views (report-code view, report-report view, and code-code view), ultimately enhancing the localization performance.

\subsection{Graph Neural Networks}

The graph data is a type of unstructured data that consists of nodes and edges, which can accurately describe the complex interaction between nodes. In recent years, graph neural networks (GNNs) have established a deep learning framework for graph structure, which can utilize both topology structure information and node feature information. 
The majorities of typical GNNs, such as graph convolutional networks (GCNs) \cite{kipf2016semi, wu2019simplifying, chen2020simple}, graph attention networks (GAT) \cite{velivckovic2017graph}, graphsage \cite{hamilton2017inductive}, etc, are designed for homogeneous graphs, which disregard the variations in both the node type and edge type.
In the real world, there is more than one kind of interaction between things, that is, there are multiple types of nodes and their varying edges. To model the heterogeneity of graphs, there are two types of approaches. The first type is to convert a heterogeneous graph into multiple homogeneous graphs, employ typical GNNs to aggregate messages, and then fused these information via attention mechanism \cite{fan2022heterogeneous}. The other type intends to model the heterogeneity directly, via different kinds of nodes and relations, represented by R-GCN \cite{schlichtkrull2018modeling}. R-GCN models the relations in graphs by employing specialized parameter matrices, which separately constructs GNNs on each relation graph.
These GNN-based models have been widely used in node classification \cite{zeng2021gcn2defect, zhou2022software}, link prediction \cite{zhang2023iea}, knowledge graphs \cite{park2019estimating} and recommender systems \cite{xie2021devrec, you2019attributed, song2024xgcn}. 

For fault localization, although there are some researches \cite{chen2008implicit, huo2020control, zhang2020exploiting, han2023bjxnet} have built localization models from the perspective of graph, few studies have considered the multiple interaction views between bug reports and source code files simultaneously. Inspired by the high accuracy and universality of graph neural networks in node-level, edge-level and graph-level analytical tasks \cite{wu2022graph}, we develop a multi-view embedded fault localization approach based on GNNs to locate the bug-relevant files.

\subsection{Contrastive Learning}

Contrastive learning is a discriminant representation learning framework, which aims to learn high-quality representation via a self-supervised manner. The common motivation behind these work is the InfoMax principle \cite{linsker1988self}, which we here instantiate as maximizing the mutual information (MI) between two views (i.e., report-code and report-report views, or report-code and code-code views). 
It learns discriminative representations through a comparative analysis of samples, distinguishing between those that are similar (positive samples) and those that are dissimilar (negative samples). This contrastive process ensures that representations of similar samples are pulled closer together in the representation space, while representations of dissimilar samples are pushed further apart, thereby enhancing the discriminative capability of the learned representations.
Contrastive learning is prevalently applied to self-supervised learning tasks, enabling the training of models on unlabeled data. Remarkably, some instances of self-supervised contrastive learning models have demonstrated performance that is comparable to supervised models. 
Due to the good performance of contrastive learning, it has been widely used in computer vision \cite{tian2020contrastive, he2020momentum} and natural language processing \cite{fu2021lrc, sun2023contrastive}. 

Inspired by the success of contrastive learning, we design an adaptive contrastive learning module to suppress noise in auxiliary information aggregation to derives more robust node representations. 
We treat the report-code bug-fixing history interaction relationships, report-report similarity relationships, and code-code co-citation relationships as different views, and maximize their mutual information to learn the collaborative signal from the auxiliary information.

\section{Problem Definitions and Preliminaries}\label{definitions}
\subsection{Motivating examples}

Let us discuss some real-world examples\footnote{The bug report examples can be found on https://bugs.eclipse.org/bugs/} that motivate our approach. 

\textbf{Scenario 1.} 

\begin{figure}[h]
	\centering
	  \includegraphics[width=0.8\linewidth]{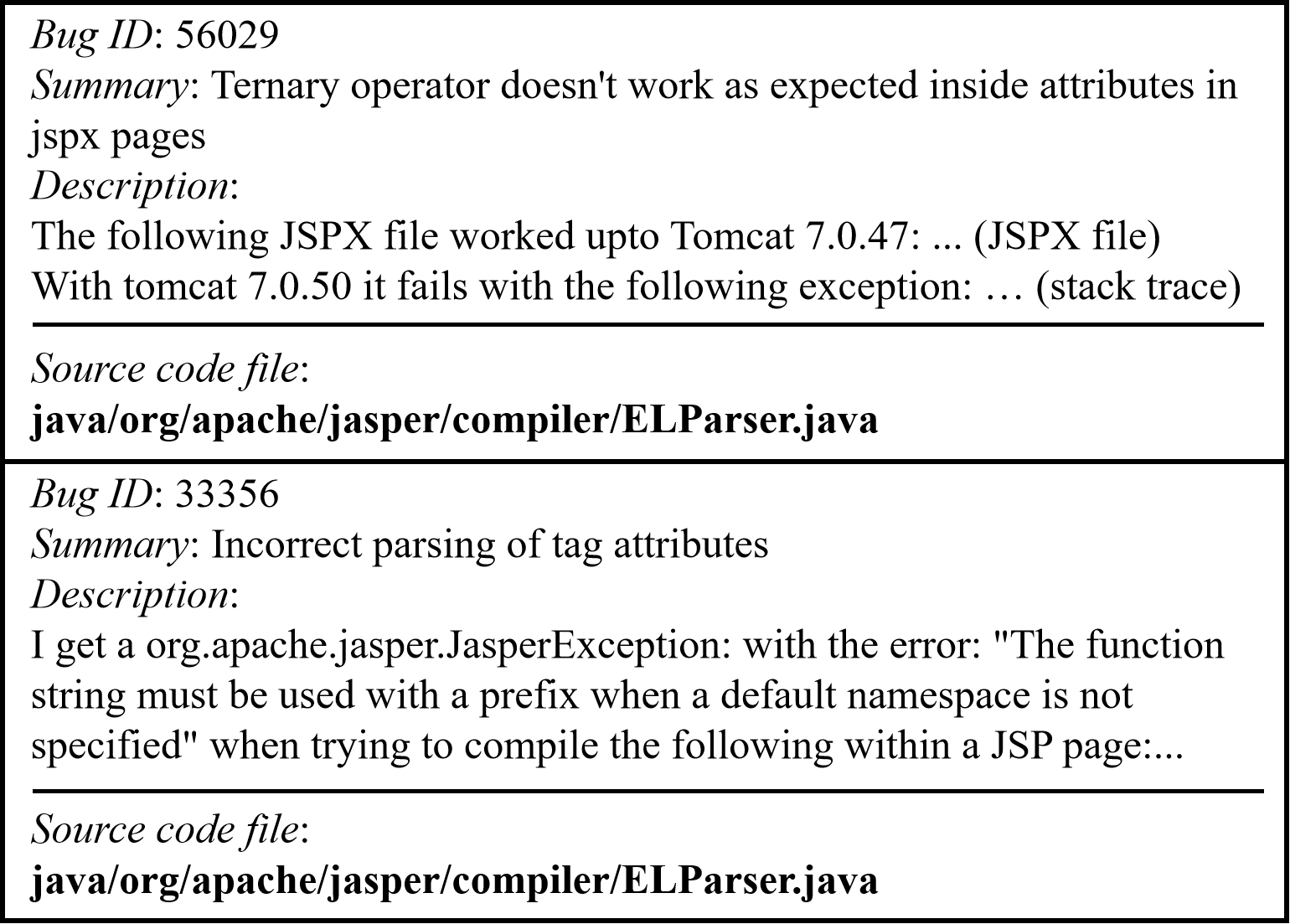}
	\caption{Bug Reports $\#56029$ and $\#33356$, and their relevant source code files.}
	\label{bug1}
\end{figure}

Fig. \ref{bug1} displays bug reports $\#56029$ and $\#33356$, and their relevant buggy files in ground truth. 

Assume that report $\#33356$ and its ground truth are known historical information, and $\#56029$ is a newly submitted bug report. Notably, an attempt to directly match the text information of the bug report $\#56029$ with candidate source code files, using cosine similarity, yielded suboptimal results, with \textit{ELParser.java} failing to rank among the top-10. 
On the one hand, the description of $\#56029$ incorporates stack trace information rich in class names, which inadvertently introduces noise into the process of matching it with relevant source code files. 
On the other hand, while the description also encompasses code snippets from JSP pages, the direct textual overlap between the front-end code in JSP pages and the back-end code in \textit{ELParser.java} is often limited. 
As a result, the combination of excessive stack trace noise and the inherent disconnect between front-end code snippets and back-end parsing logic contributed to the suboptimal ranking of \textit{ELParser.java} in the context of this specific bug report. 
From another perspective, we can see that both reports $\#33356$ and $\#56029$ describe an exceptional bug related to the parsing of tag attributes within JSP pages. By exploiting the historical knowledge embodied in report $\#33356$, which indicates that the buggy file associated with its resolution was \textit{ELParser.java}, we can hypothesize that similar bugs may also involve this file \cite{zhou2012should, davies2012using, li2021survey}. 
We leverage the potential of aggregating information from similar bug reports to indirectly infer the involvement of potentially related files, thereby improving the efficiency and accuracy of locating the root cause of bugs.

\begin{figure}[h]
	\centering
	  \includegraphics[width=\linewidth]{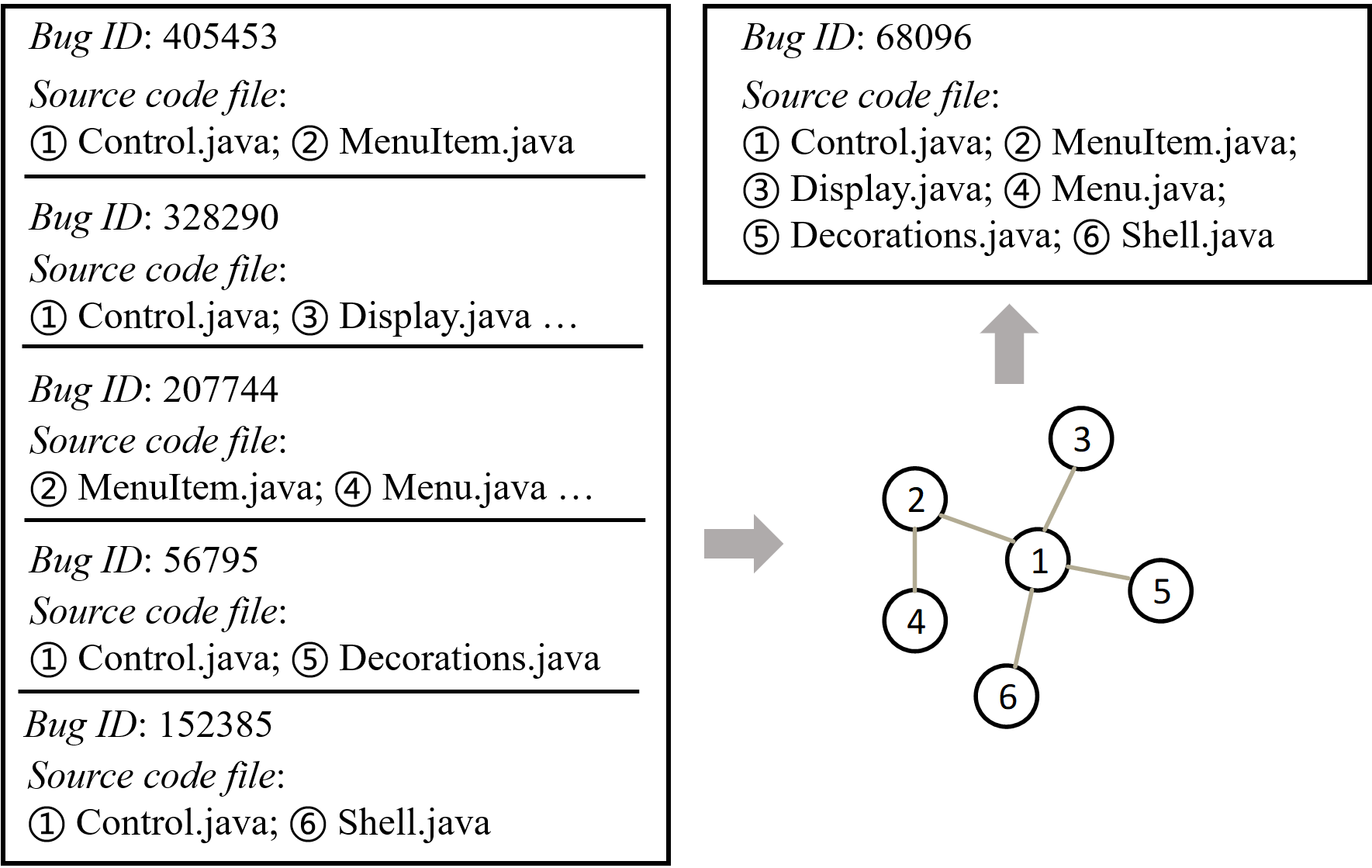}
	\caption{Bug Reports and their relevant source code files. The node number in the co-citation graph is the index of each source code file.}
	\label{bug2}
\end{figure}

\textbf{Scenario 2}. Fig. \ref{bug2} shows six bug reports and their relevant buggy files in ground truth. Assume that the five reports on the left are historical data and $\#68096$ is a newly submitted report.
If we simply match report $\#68096$ to each individual code file, it could be very hard to locate all the true buggy files together. 
However, if considering the co-citation relationships among these files in historical bug reports and build a network, we can observe the coupling connection between them.
For example, according to the bug-fixing history of the five bug reports, the generated connections include $(1,2),(1,3),(2,4),(1,5),(1,6)$. Then, a co-citation graph can be constructed with the connection link between each co-cited pair of files, as shown in Fig. \ref{bug2}. 
As a consequence, we can quickly recommend these buggy files as potential candidate locations to report $\#68096$ through the co-citation network.

Those motivating examples suggest that, while finding the source code files relevant to a bug report, the similarity relationship between bug reports and the co-citation relationship between source code files could provide auxiliary information to help better localization.

\subsection{Definitions}

In this paper, we regard fault localization as one of the recommendation scenarios. We formally introduce the following definitions before describing our proposed framework. The notations in this paper are shown in Table \ref{Notations}.

\textit{Definition 1 (Report-Code Interaction Relationship)}. Following the settings in GNN-based recommendation, we construct a report-code bipartite graph based on the bug-fixing history records, which reflects the relevance between bug reports and source code files. Let $R=\{r_1,r_2,...,r_m\}$ denotes a set of bug reports, and $C=\{c_1,c_2,...,c_n\}$ represents the current project, where $m$ is the number of bug reports in a period, and $n$ is the number of code files in this project.
We construct report-code interaction matrix $\mathbf{A}_{R-C}=\{a_{rc}(r_i,c_j)|r \in R,c \in C\}$, where

\begin{equation}
\label{Eq1}
a_{rc}(r_i,c_j)=
\begin{cases}
tex\_sim(r_i,c_j)& \text{if edge} <r_i,c_j> \text{exists,}\\
0& \text{otherwise}
\end{cases}
\end{equation}
where $tex\_sim(r_i,c_j) = \frac{\mathbf{e}_{r_i} \cdot \mathbf{e}_{c_j}}{||\mathbf{e}_{r_i}||||\mathbf{e}_{c_j}||}$ is use to calculate a similarity score between the report  $r_i$ and the code file $c_j$, which serves as the weight of the edge  $<r_i,c_j>$. Here, $\mathbf{e}_{r_i}$ and $\mathbf{e}_{c_j}$ represent the vector representations of  $r_i$ and $c_j$ respectively, which are obtained by embedding these entities using CodeBERT \cite{feng2020codebert}.

\textit{Definition 2 (Report-Report Similarity Relationship)}. Similar bug reports might correspond to similar buggy modules. Based on this assumption, we construct a report-report graph, and the adjacency matrix is $\mathbf{A}_{R-R}=\{a_{rr}(r_i,r_j)|r \in R\}$, where

\begin{equation}
\label{Eq2}
a_{rr}(r_i,r_j)=
\begin{cases}
tex\_sim(r_i,r_j)& \text{if edge} <r_i,r_j> \text{exists,}\\
0& \text{otherwise}
\end{cases}
\end{equation}
each value of $tex\_sim(r_i,r_j) = \frac{\mathbf{e}_{r_i} \cdot \mathbf{e}_{r_j}}{||\mathbf{e}_{r_i}||||\mathbf{e}_{r_j}||}$ is the weight of edge $<r_i,r_j>$, which indicates the cosine similarity score of bug reports $r_i$ and $r_j$. 

\textit{Definition 3 (Code-Code Co-citation Relationship)}. The implicit relationships between fault locations that are co-cited by the same bug reports can be further modeled as a co-citation network to help predict faults \cite{chen2008implicit}. As such, we construct a code-code graph, and the adjacency matrix $\mathbf{A}_{C-C}=\{a_{cc}(c_i,c_j)|c \in C\}$, where

\begin{equation}
\label{Eq3}
a_{cc}(c_i,c_j)=
\begin{cases}
co\_cited\_count(c_i,c_j)& \text{if edge} <c_i, c_j> \text{exists,}\\
0& \text{otherwise}
\end{cases}
\end{equation}
each value of $co\_cited\_count(c_i,c_j)$ represents the number of times that the code files $c_i$ and $c_j$ were fixed by the same bug report, which serves as the weight of the edge $<c_i, c_j>$.

\textit{Definition 4 (Recommendation Task)}. We construct the entire dataset as three views (report-code interaction view,  report-report similarity view and code-code co-citation view). The goal of our recommendation task is to learn a function $\hat{y}(r,c)$ based on these views that can predict the probability that report $r$ will fix with code file $c$.

\begin{table}[h]
\caption{Notations utilized in the paper.}
  \label{Notations}
\begin{tabular*}{\textwidth}{@{\extracolsep\fill}cl}
\toprule
Notations & Description \\
\midrule
$R$ & $R=\{r_1,r_2,...,r_m\}$ denotes a set of bug reports. \\
$C$ & $C=\{c_1,c_2,...,c_n\}$ denotes a set of code files. \\
$\mathbf{A}_{R-C}$ & report-code matrix based on the interaction relationship between bug \\ & reports and code files. \\
$\mathbf{A}_{R-R}$ & report-report matrix based on the similarity relationship between bug \\ & reports. \\
$\mathbf{A}_{C-C}$ & code-code matrix based on the co-citation relationship between code files. \\
$a_{rc}(r_i,c_j)$ & the interaction relationship between $r_i$ and $c_j$. \\
$a_{rr}(r_i,r_j)$ & the similarity relationship between $r_i$ and $r_j$. \\
$a_{cc}(c_i,c_j)$ & the co-citation relationship between $c_i$ and $c_j$. \\
$\mathbf{e}_v$ & the embedding vectors of node $v$. \\
\midrule
\multicolumn{2}{l}{Notations in Section \ref{proposed framework}} \\
\midrule
$G_{R-C}, G_{R-R}, G_{C-C}$ & denote the report-code graph, report-report graph, and code-code graph, \\ & respectively. \\
$N_v^{L=i}$ & the neighbor set of node $v$ at layer $i$. \\
$\mathbf{e}_r$ & the embedding vectors of node $r$ in report-report view after GNN \\ & embedding layer. \\
$\mathbf{e}_c$ & the embedding vectors of node $c$ in code-code view after GNN embedding \\ & layer. \\
$\mathbf{h}_r, \mathbf{h}_c$ & the embedding vectors of node $r$ or $c$ in report-code view after GNN \\ & embedding layer. \\
$\mathbf{z}_v^p, \mathbf{z}_v^q$ & the mapped embedding vectors of node $v$ after contrastive learning. \\
\bottomrule                
\end{tabular*}
\end{table}

\section{Proposed Framework}\label{proposed framework}

\begin{figure}
	\centering
	  \includegraphics[width=\textwidth]{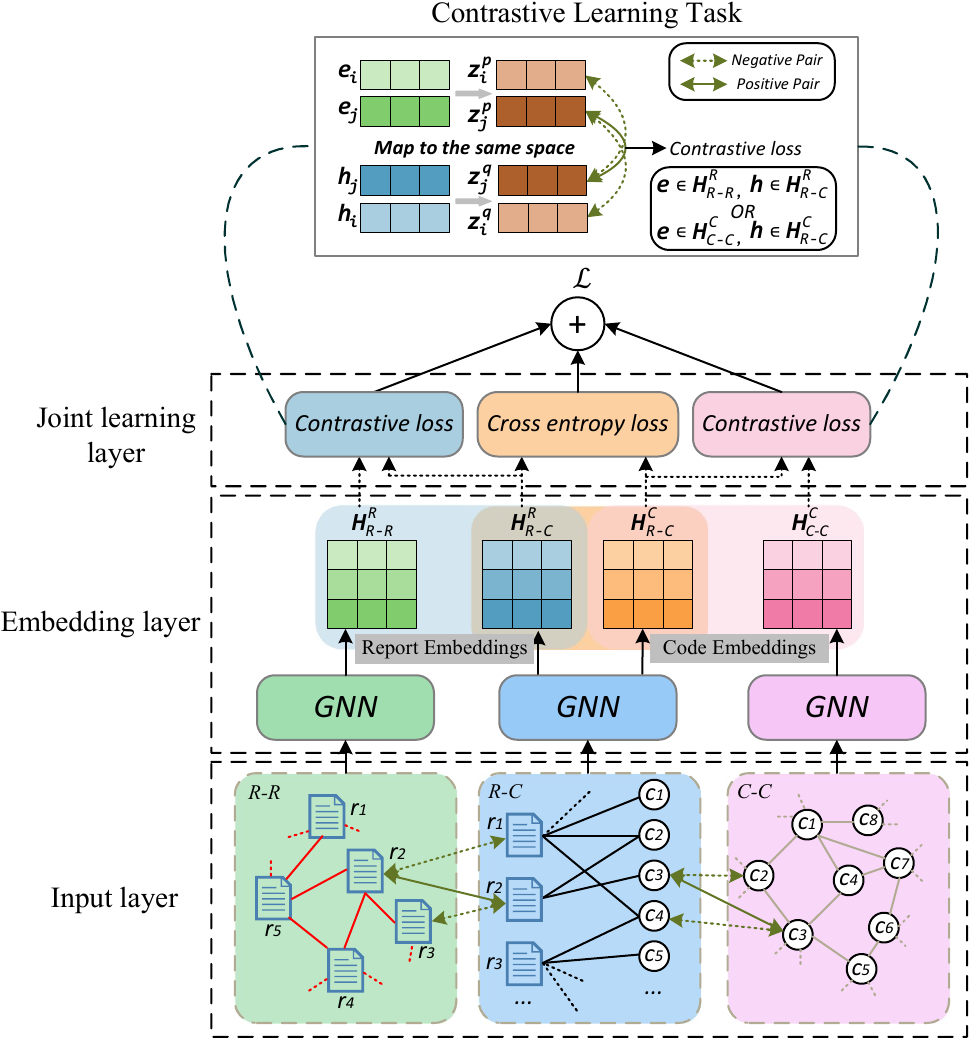}
	\caption{The main framework. Where R-C denotes the report-code interaction view, R-R denotes the report-report similarity view, and C-C denotes the code-code co-citation view. $\mathbf{H}_{R-R}^R$ and $\mathbf{H}_{R-C}^R$ represent the bug report representation matrix updated by report-report view and report-code view after GNN embedding layer. While $\mathbf{H}_{C-C}^C$ and $\mathbf{H}_{R-C}^C$ represent the source code representation matrix updated by code-code view and report-code view after GNN embedding layer.}
	\label{framework}
\end{figure}

In this section, we propose the Multi-View Adaptive Contrastive Learning  framework (MACL-IRFL), for recommending top-$N$ potential bug-relevant files to each bug report. 
We explicitly encode the \textit{Report-Code Interaction Relationship}, \textit{Report-Report Similarity Relationship} and \textit{Code-Code Co-citation Relationship}
into the representation learnings of bug reports and code files, and the architecture is illustrated in Fig. \ref{framework}. The framework includes three components: 1) the input layer, which inputs the three views (report-code interaction view,  report-report similarity view and code-code co-citation view) and initializes the embeddings of bug reports and code files; 
2) the embedding aggregation layer, which iteratively aggregates attributed neighbor node vectors around each node, and updates the embedding of the nodes by stacking multiple layers; 3) the joint learning layer, which combines contrastive learning task to encode the information shared between report-report and report-code views (or code-code and report-code views) for learning high-quality representations. 

\subsection{Input Layer}

We first construct the report-code, report-report and code-code graphs based on the definitions in Section \ref{definitions}. 
(1) Report-Code graph. Researchers could collect a large number of bug reports and bug-relevant files during the life cycle of a software project. Based on the explicit interaction relationships between bug reports and bug-relevant files, we construct a report-code bipartite graph $G_{R-C} = (V_{R-C}, E_{R-C})$, where $V_{R-C}(|V_{R-C}|=|R|+|C|=m+n)$ denotes the node set, $m$ and $n$ denote the total number of bug reports and code files, respectively. $E_{R-C}$ denotes the edge set. 
(2) Report-Report graph.
We first preprocess the texts in reports (e.g., removing the stop words, symbols and spaces), and then convert the processed texts into vector representations using CodeBERT. Next, we construct a report-report graph $G_{R-R} = (V_{R-R}, E_{R-R})$, where $V_{R-R}(|V_{R-R}|=|R|)$ denotes the report node set, and $E_{R-R}$ denotes the similarity edge set.
(3) Code-Code graph. Fig. \ref{example}(b) illustrates the process of the co-citation graph construction. For example, there are three bug reports $r_1$, $r_2$, and $r_3$, and their relevant code files. The code files $c_1,c_2,c_4$ are co-cited by bug report $r_1$ and can be further modeled as a small co-citation network. Similarly, the relevant code files of $r_2$, and $r_3$ are modeled as a co-citation network respectively. Finally, a complete code-code graph is formed as $G_{C-C} = (V_{C-C}, E_{C-C})$, where $V_{C-C}(|V_{C-C}|=|C|)$ denotes the code file node set, and $E_{C-C}$ denotes the co-citation edge set.

\subsection{GNN Embedding Layer}
\begin{figure}
	\centering
	  \includegraphics[width=0.8\textwidth]{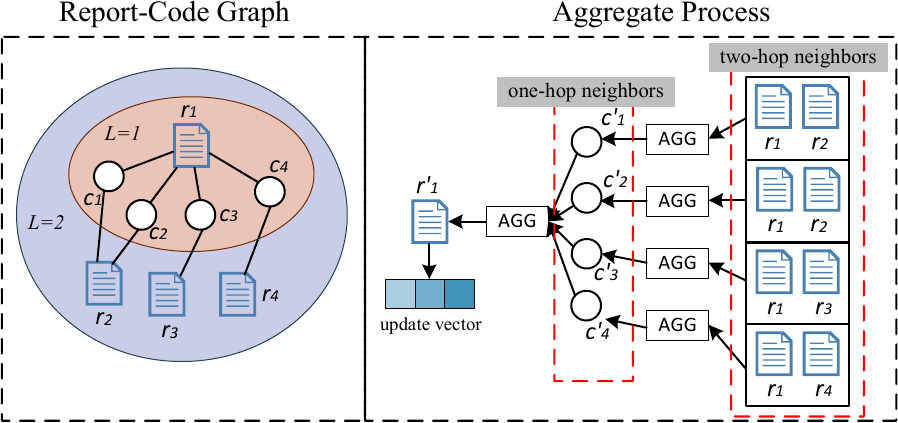}
	\caption{The aggregation process of GNN embedding layer. We take report-code graph with two-layer aggregation as an example. $r_i$ and $c_i$ represent the initial node, while $r_i^{'}$ and $c_i^{'}$ represent the updated nodes. }
	\label{aggregate_process}
\end{figure}
In this section, we explicitly encode the report-code interaction relationship, report-report similarity relationship, and code-code co-citation relationship into the representation learnings of bug reports and code files based on the architecture of GAT. We start by describing a two-layer of GAT in report-code graph as an example, which includes the process of neighborhood aggregation and representation updating. As shown in  Fig. \ref{aggregate_process}, considering a report node $r_1$, we use $N^{L=1}_{r_1} = \{c_1,c_2,c_3,c_4\}$ to represent the one-hop neighbors of $r_1$. The neighboring nodes of $\{c_1,c_2,c_3,c_4\}$ are 
$N^{L=1}_{c_1} = \{r_1,r_2\}$, $N^{L=1}_{c_2} = \{r_1,r_2\}$, $N^{L=1}_{c_3} = \{r_1,r_3\}$, and $N^{L=1}_{c_4} = \{r_1,r_4\}$.
That is, the two-hop neighbors of $r_1$ is $N^{L=2}_{r_1} = \{r_1,r_2,r_3,r_4\}$. In aggregation process, the neighbor nodes in $N^{L=2}_{r_1}$ propagate their information to the node in $N^{L=1}_{r_1}$. For example, $r_1,r_3$ propagate the initial representation to $c_3$, and then the representation of $c_3$ is updated. After that, the node feature vectors in $N^{L=1}_{r_1}$ are updated and propagated the new message to $r_1$. Finally, the updated representation of $r_1$ is the final two-hop output. 
This higher-order aggregation mechanism enables $r_1$ not only to learn the details of its related code files but also indirectly acquire information from other reports that are connected to those code files or share similar contexts or attributes.
The nodes of report-report graph and code-code graph are similar to the above steps.

\subsection{Joint Learning Layer}
In the joint learning layer, the objective task comprises two parts. The first is our main task, i.e., fault localization task. After performing the GNN embedding aggregation stage, we can obtain the updated representations for bug report $r$ and code file $c$ from report-code graph. We calculate the inner product of the representations of bug report $r$ and code file $c$ to predict their matching score as:
\begin{equation}
\label{Eq4}
\hat{y}_{R-C}(r,c)=z_r^T \cdot z_c
\end{equation}

The fault localization task is considered as a ranking problem and the loss function we adopt the pointwise Cross-Entropy loss \cite{mao2023cross} during the training process, which can compute a probability distribution over all candidate code files and maximize the difference between the scores of positive samples and negative samples as much as possible. For report-code interaction relationship, the observed interactions should be given a higher prediction score than those unobserved interactions. Therefore, the objective function as follows:

\begin{equation}
\label{Eq5}
L_{FL} = \sum_{(r,c) \in O} -\log \frac{exp(\hat{y}_{R-C}(r,c))}{ \sum^{|C|}_{j=1} exp(\hat{y}_{R-C}(r,j))}
\end{equation}
where $\forall (r,c) \in O$ denotes the set of observed interactions, i.e., positive instances. $|C|$ is the total number of candidate code files.

Note that the representation space of report-code, report-report and code-code views are different. Therefore, the second task is to leverage contrastive learning to encourage graph encoders to preserve the information shared across report-code interaction view to report-report and code-code views.
As shown in Fig. \ref{framework}, we feed report embeddings $(\mathbf{e}_r, \mathbf{h}_r)$ from report-report view and report-code view into two correspond MLPs, mapping them into the same space $(\mathbf{z}^p_r, \mathbf{z}^q_r)$, where contrastive loss of report representations is calculated. Similarly, code embeddings $(\mathbf{e}_c, \mathbf{h}_c)$ from code-code view and report-code view are mapped into $(\mathbf{z}^p_c, \mathbf{z}^q_c)$. 
In order to pull the positive samples closer together in the representation space and push the negative samples further apart,
for each report $r_i$, we treat the report-report view and report-code view of the same report as a positive pair $\{\mathbf{z}^p_{r_i}, \mathbf{z}^q_{r_i}\}$. On the other side, we pair report $r_i$ with a random report $r_j$, and get $\{\mathbf{z}^p_{r_i}, \mathbf{z}^q_{r_j}\}$ and $\{\mathbf{z}^p_{r_j}, \mathbf{z}^q_{r_i}\}$ as negative pairs. Then the contrastive loss function promotes the consistency between the representations of the two different views, while maximizes the divergence of negative pairs. That is, $sim(\mathbf{z}^p_{r_i}, \mathbf{z}^q_{r_i}) > sim(\mathbf{z}^p_{r_i}, \mathbf{z}^q_{r_j})$ and $sim(\mathbf{z}^p_{r_i}, \mathbf{z}^q_{r_i}) > sim(\mathbf{z}^p_{r_j}, \mathbf{z}^q_{r_i})$. 
Analogously, the contrastive learning process for code representations follows the above steps, emphasizing consistency and divergence across code-code view and report-code view. Formally, we adopt the contrastive learning object function based on InfoNCE loss \cite{oord2018representation}:

\begin{equation}
\label{Eq8}
L_{CL} = -(\log \frac{exp(sim(\mathbf{z}^p_{r_i}, \mathbf{z}^q_{r_i}) / \tau)}{ \sum^{|R|}_{j=1, j \neq i} exp(sim(\mathbf{z}^p_{r_i}, \mathbf{z}^q_{r_j}) / \tau)} + \log \frac{exp(sim(\mathbf{z}^p_{c_i}, \mathbf{z}^q_{c_i}) / \tau)}{ \sum^{|C|}_{j=1, j \neq i} exp(sim(\mathbf{z}^p_{c_i}, \mathbf{z}^q_{c_j}) / \tau)})
\end{equation}
where $sim(\mathbf{z}^p_{i}, \mathbf{z}^q_{j}) = \frac{\mathbf{z}^p_{i} \cdot \mathbf{z}^q_{j}}{||\mathbf{z}^p_{i}||||\mathbf{z}^q_{j}||}$ measures the cosine similarity of two vectors, $|R|$ and $|C|$ are the total number of bug reports and code files. $\tau$ denotes the temperature hyper-parameter.

Finally, we leverage a multi-task training strategy to jointly optimize the fault localization loss and the contrastive learning loss. The overall loss function is:
\begin{equation}
\label{Eq7}
L = L_{FL} + \lambda L_{CL}
\end{equation}
where $\lambda$ is the hyper-parameter to balance different terms. In practice, $\lambda$ is fixed as 0.01.

\subsection{Model Prediction}
As mentioned above, to optimize our objective task, we decouple the training process into two iterative stages: fault localization task and contrastive learning task. We iteratively update the corresponding parameters to minimize the losses until we reach the best performance on the validation set. In particular, we update the parameters of report-code view in the fault localization loss, and then freeze the parameters in the contrastive learning to adaptively filter out unimportant edges. During the testing phase, where new bug reports lack fixing history and thus the direct report-code interaction view is unavailable. We emphasize that the representations in test stage rely on the report-report and code-code views and infer relationships between the new bug reports and code files. Leveraging the alignment achieved during the training phase through contrastive learning between the two views (i.e., report-report and code-code views) and the report-code view, we update the representations within the report-report and code-code views using the learned parameters. Subsequently, we calculate similarity scores between each report and code file, which are then sorted to recommend the top-$N$ code files most relevant to the given bug report. 

\section{Experimental results and analysis}\label{experiments}

In this section, we first describe the experimental setup, including the introduction of dataset, evaluation metrics, baselines, implementation details. And then we present the main experimental results and our analysis.

\subsection{Experimental setup}
\subsubsection{Dataset}

To evaluate the effectiveness of our proposed approach, we use the same benchmark datasets provided by Ye et al. \cite{ye2014learning}, which are widely used in fault localization research works \cite{zhou2012should, ye2016word, lam2017bug, gharibi2018leveraging, xiao2017improving, xiao2019improving, zhang2020exploiting, qi2021dreamloc, liang2022modeling, zhu2022bl, han2023bjxnet}. 
The datasets have five real-world projects (\textit{Birt}, \textit{Eclipse Platform UI}, \textit{JDT}, \textit{SWT}, \textit{Tomcat}), which contain ground truth associated with each bug report. 
For comparison, we follow previous studies to evaluate our approach to each project separately.

Table \ref{dataset} shows the statistics of the datasets. The column of \textit{\#Bug Reports} shows the number of bug reports of each project. 
In line with Qi et al. \cite{qi2021dreamloc}, since some buggy files may be modified or deleted in the fixed version, for each bug we check out the source code files of the prefix version, and remove the invalid buggy paths which do not exist in the prefix version. After that, we filter out bug reports that without linking any buggy paths. 
Furthermore, we filter out bug reports from the validation and test sets whose ground truth (i.e., relevant buggy files) are not present in the training set. 
The values in parentheses are the number of bug reports we filtered. 
The column of \textit{\#Source Files} shows the number of source code file type nodes in report-code graph and code-code graph. Specifically, we chronologically sort bug reports by commit timestamp and choose the oldest version as the first version. We keep all the source code files from the first version as nodes in the network. Since the number of source code files in the next version may be changed by adding or deleting files, we perform the following steps: 1) for newly added files, we directly add them to the network; 2) for deleted files, we do not remove them from the network, since they may be linked to bug reports from the previous version; 3) for modified files, if only the contents of the file are modified, the nodes in the network do not make any changes. If the name of the modified file is different from the previous version, we add the modified file as a new node to the network. Follow these steps until the latest version in training set has been traversed, and the final number of source code files of each project is shown in column \textit{\#Source Files}.
The column of \textit{\#Buggy Files} shows the number of fixed buggy files, and \textit{\#Buggy Rate} indicates the percentage of buggy files within each project.

Table \ref{dataset-interactions} shows the statistics of the three types of relationships. The second to fourth columns show the number of report-code interaction relationships, report-report similarity relationships, and code-code co-citation relationships, respectively. The fifth column is the average number of fixed buggy files of each bug report, and the sixth column is the average number of similar reports of each bug report. 
When constructing co-citation relationships, we filter out those relationships that commit more than 10 files at a time, as such \textit{large commits} are often attributed to generic modifications such as grammatical corrections, branch merges, and other non-specific changes, resulting in weaker connections between files and thus constituting noise data.
Since the co-citation relationships only exist between the co-cited buggy files by the same report, we only count the average number of co-citation relationships of each buggy file (shown in the last column of Table \ref{dataset-interactions}), not for all source code files.

As we can see, the five projects are of different scales. \textit{Eclipse Platform UI} and \textit{JDT} have the largest number of bug reports, about six thousand reports.
\textit{Birt} and \textit{SWT} have a medium size of bug reports, about four thousand reports. And \textit{Tomcat} has the small size, only around one thousand reports.

\begin{table}[h]
\caption{Statistics of benchmark dataset. $\#$Bug Reports represents the number of bug reports of each project, and the values in parentheses represent the actual number of bug reports filtered in this paper.}
  \label{dataset}
\begin{tabular*}{\textwidth}{@{\extracolsep\fill}cccccc}
\toprule
Project & Time Range & $\#$Bug Reports & $\#$Source Files & $\#$Buggy Files & $\#$Buggy Rate \\
\midrule
Birt    & 06/2005-12/2013 & 4,178 (4,009) & 8,799 & 2,468 & 28.05\% \\
Eclipse* & 10/2001-01/2014 & 6,495 (6,261) & 13,327 & 2,068 & 15.52\% \\
JDT        & 10/2001-01/2014 & 6,274 (6,108) & 14,167 & 1,892 & 13.35\% \\
SWT        & 02/2002-01/2014 & 4,151 (4,086) & 3,587 & 768 & 21.41\% \\
Tomcat     & 07/2002-01/2014 & 1,056 (1,020) & 2,755 & 625 & 22.69\% \\
\bottomrule
\multicolumn{4}{l}{\small Eclipse* refers to Eclipse Platform UI.}\\ 
\end{tabular*}
\end{table}

\begin{table}[h]
\caption{Statistics of three types of interactions of each project. $\#$RC, $\#$RR and $\#$CC denote the number of report-code relationships, report-report relationships and code-code relationships, respectively.}
  \label{dataset-interactions}
\begin{tabular*}{\textwidth}{@{\extracolsep\fill}ccccccc}
\toprule
\multirow{2}{*}{Project} & \multirow{2}{*}{$\#$RC} & \multirow{2}{*}{$\#$RR} & \multirow{2}{*}{$\#$CC} & \multicolumn{1}{c}{Avg.$\#$RC per} & \multicolumn{1}{c}{Avg.$\#$RR per} & \multicolumn{1}{c}{Avg.$\#$CC per} \\ 
& & & & Bug Report & Bug Report & Buggy File \\
\midrule
Birt    & 7,880 & 35,525 & 12,254 & 1.97 & 8.86 & 4.97 \\
Eclipse* & 8,455 & 104,756 & 15,552 & 1.35 & 16.73 & 7.52 \\
JDT        & 9,608 & 82,450 & 15,238 & 1.57 & 13.50 & 8.05 \\
SWT        & 6,697 & 150,964 & 4,984 & 1.64 & 36.95 & 6.49 \\
Tomcat     & 1,647 & 4,911 & 1,940 & 1.61 & 4.81 & 3.10  \\
\bottomrule
\end{tabular*}
\end{table}

\subsubsection{Evaluation metrics}
To evaluate our approach, we use Accuracy@N, mean average precision (MAP) \cite{schutze2008introduction}, and mean reciprocal rank (MRR) \cite{voorhees2001trec} as the evaluation metrics, which are commonly used in software engineering. The higher the Accuracy@N, MRR, and MAP values, the better the fault localization performance.

Accuracy@N is the percentage of bug reports whose at least one buggy file is ranked in the top $N$ files in the recommended list. It measures the overall ranking performance of the localization model, which is broadly used in prior studies \cite{qi2021dreamloc}. We consider three values of $N$: 1, 5, and 10.

MAP is a metric to evaluate the average performance of locating all relevant files, which is suitable for evaluating the ranking quality of a method when a bug involves multiple buggy files. Supposing that the relevant buggy files of $i$-th bug report are retrieved and ranked as $S=\{s_1,s_2,...,s_{|relevant|}\}$. AP and MAP are computed as follows:

\begin{align}
MAP = \frac{1}{|R|} \sum_{i=1}^{|R|} AP(i) \\
AP(i) = \sum_{j=1}^{M} \frac{(T(j)/j) \times pos(j)}{|relevant|} \label{Eq9}
\end{align}
where $M$ is the maximum position of $S$ for the $i$-th bug report, and $|R|$ is the number of bug reports. $pos(j)$ returns 1 if the $j$-th buggy file in the list is correct and 0 otherwise. $T(j)$ denotes the number of buggy files in the top $j$.

MRR measures how well the first buggy file is correctly located and ranked in the recommended file list, which is formulated as:
\begin{equation}
\label{Eq11}
MRR = \frac{1}{|R|} \sum_{i=1}^{|R|} \frac{1}{first_i}
\end{equation}
where $first_i$ denotes the position of the first correctly-located buggy file for the $i$-th bug report.

Moreover, we use the Wilcoxon signed rank test \cite{wilcoxon1970critical} to check whether the performance difference between our model with baselines is significant. We consider that one approach performs significantly better than the other one at the 95\% confidence level if the corresponding $p$-value is less than 0.05, otherwise, there is no statistically significant difference between the two approaches. 
We also employ Cliff's delta ($\delta$) \cite{macbeth2011cliff} to quantify the amount of difference between two approaches. Here, the effectiveness level is considered negligible if $|\delta| < 0.147$, small if $0.147 \leq |\delta| < 0.33$, medium if $0.33 \leq |\delta| < 0.474$, and large if $|\delta| \geq 0.474$.

\subsubsection{Baselines}\label{baselines}
To measure the relative effectiveness of our model, we compare our model against the following representative approaches.

1) BugLocator \cite{zhou2012should}: an IR-based approach that utilized revised vector space model (rVSM) to calculate the text similarity between bug reports and source code files, and then ranked all the source code files to locate potential buggy files to a given bug report.

2) DNNLOC \cite{lam2017bug}: a deep learning approach that combined the features built from deep neural network (DNN), rVSM, and project's bug-fixing history. 

3) KGBugLocator \cite{zhang2020exploiting}: a knowledge graph and deep learning approach that constructed code knowledge graph to enrich semantic information of code for localization.

4) DreamLoc \cite{qi2021dreamloc}: a deep learning approach that employed the attention mechanism to calculate the relevance between bug report terms and code snippets.

5) FLIM \cite{liang2022modeling}: a deep learning approach that utilized a large language model (i.e., CodeBERT) to extract semantic features of source code file at the function-level.

6) BL-GAN \cite{zhu2022bl}: a semi-supervised learning approach that employed generative adversarial network to capture the potential relevance distribution between bug report and source code file from both limited bug-fix records and abundant unfixed bug reports.

7) bjXnet \cite{han2023bjxnet}: a deep learning approach that combined the text semantic features of the source code and the graph features of the code property graph to enrich semantic information of code for localization.

\subsubsection{Implementation details}
We use pytorch to construct the model and all experiments are run on Windows Server 2016 Datacenter with 6 cores of 2.3 GHz CPU, 48 GB RAM and NVIDIA Tesla V100 GPUs with 32 GB memory. 
We use the uniform distributed random initializer to generate the initial embeddings of bug reports and source code files.
GNN training is at full batch, and the embedding size defaults to 32 for each node in the three views.
Here, we optimize our model with the Adam optimizer \cite{kingma2014adam} with learning rate 0.01 for training.
For regularization, we apply the early stopping strategy \cite{caruana2000overfitting} for training.
In all experiments, the optimal number of GNN layers is searched in $\{1, 2, 3, 4, 5\}$, which is discussed in detail in RQ3 of Section \ref{RQs}. For the projects \textit{Birt}, \textit{JDT}, and \textit{Tomcat}, the number of layers is set to 4, while for \textit{Eclipse Platform UI} and \textit{SWT}, is set to 3.
The temperature of contrastive learning $\tau$ is set to 0.1.
To avoid over-fitting, we dropout (dropout rate is 0.2) in every layer of our proposed MACL-IRFL.
We split each dataset into a training dataset containing 80\% data, a validation dataset containing 10\% and a test dataset containing 10\% data by chronological order.

\subsection{Research Questions}\label{RQs}
In the experiments, we seek to answer the following research questions:

\textbf{RQ1: How does the proposed model perform compared to the baselines?}

To answer RQ1, we choose seven baseline models for comparison: BugLocator, DNNLOC, KGBugLocator, DreamLoc, FLIM, BL-GAN, and bjXnet. BugLocator is an IR-based model, whose core idea is to match textual similarities between source code files and bug reports to locate buggy files. DNNLOC is a deep learning model that use DNN to help bridge the lexical gap by learning to link high-level, abstract concepts between bug report and source file. 
KGBugLocator and bjXnet enrich the semantic information by building the source code into code knowledge graph and CPG respectively.
DreamLoc and FLIM extract semantic information by splitting the source code into code snippets and functions, respectively. BL-GAN improves localization performance from the perspective of alleviating limited bug-fix records and abundant unfixed bug reports.
We test our model on five open-source projects and compare it with these baseline approaches. 
Since these approaches use the same projects and metrics to evaluate the performance of the proposed tools, we use the results reported in their papers for comparison.
However, KGBugLocator hasn't reported its results for \textit{Brit} and \textit{Eclipse Platform UI}, and bjXnet hasn't reported its results for \textit{Brit} and \textit{Tomcat}. BL-GAN also hasn't reported its results for \textit{Brit} and \textit{SWT} and there are no MRR results for the other four projects. Therefore, we can not compare our results against their approaches for these projects.

\textbf{RQ2: How do adaptive contrastive learning strategy impact our model?}

To answer RQ2, we design ablation experiments to investigate whether our proposed model can benefit from the adaptive contrastive learning strategy. 
We construct three variants of our MACL-IRFL as shown in Table \ref{variants}, $variant_0$ can only access the report-code view in training phase. $variant_1$ incorporates both report-report and report-code views in training phase, enabling adaptive sharing of report representations across these two views. And $variant_2$ incorporates both code-code and report-code views in training phase, enabling adaptive sharing of code representations across these two views. 
Our model, MACL-IRFL, is capable of simultaneously engaging in adaptive sharing of report representations across both the report-report and report-code views, as well as sharing code representations between the code-code and report-code views.
During the validation and testing phases, all of these variants employ the report-report and code-code views to learn the report and code representations respectively.

\begin{table}[h]
\caption{Construct the different variants of our model.}
  \label{variants}
\begin{tabular*}{\textwidth}{@{\extracolsep\fill}cl}
\toprule
\multicolumn{2}{l}{The variants descriptions} \\
\midrule
$variant_0$ & Remove the report-report and code-code views, and only consider the \\ & report-code interaction relationships in training phase. \\
\midrule
$variant_1$ & Remove the code-code view, and only consider the report-code interaction \\ & relationships and report-report similarity relationships in training phase. \\
\midrule
$variant_2$ & Remove the report-report view, and only consider the report-code interaction \\ & relationships and code-code co-citation relationships in training phase. \\
\midrule
MACL-IRFL & The model in this paper, considers the report-code interaction relationships, \\ & code-code co-citation relationships and report-report similarity relationships \\ & in training phase. \\
\bottomrule                
\end{tabular*}
\end{table}

\textbf{RQ3: Does the embedding propagation from GNN high-layer help improve the localization performance?}

Same as RQ2, we conduct ablation experiments to investigate the impact of aggregating high-order neighbors on our model. Previous approaches use bug-fix records to mine the relevance between bug reports and source code files, which maps to our graphs indicating that each node aggregates its direct one-hop neighbors information. 
Here, we aim to evaluate the benefits of using high-order connectivity on localization performance. We vary the number of GNN layers $L$ of our model from 1 to 5 and compare their performance under different layers.

\subsection{Experimental Results for Research Questions}
\textbf{RQ1: How does the proposed model perform compared to the baselines?}

\begin{table}[h]
\caption{Detailed comparison results of eight approaches.}
\label{RQ1-performance-tab}
\begin{tabular*}{\textwidth}{@{\extracolsep\fill}ccccccc}
\toprule
\multirow{2}{*}{Project} & \multirow{2}{*}{Model} & \multicolumn{3}{c}{Accuracy@N} & \multirow{2}{*}{MAP} & \multirow{2}{*}{MRR} \\\cmidrule{3-5} 
& & n=1 & n=5 & n=10 & & \\
\midrule
\multirow{8}{*}{Birt} 
& BugLocator & 0.11 & 0.25 & 0.32 & 0.14 & 0.18 \\
& DNNLOC & 0.25 & 0.42 & 0.51 & 0.20 & 0.28 \\
& KGBugLocator & - & - & - & - & - \\
& Dreamloc & 0.31 & 0.56 & 0.65 & 0.36 & 0.42 \\
& FLIM & 0.18 & 0.36 & 0.45 & 0.21 & 0.27 \\
& BL-GAN & - & - & - & - & - \\
& bjXnet & - & - & - & - & - \\
& MACL-IRFL & \textbf{0.56} & \textbf{0.80} & \textbf{0.87} & \textbf{0.59} & \textbf{0.67} \\
\midrule
\multirow{8}{*}{Eclipse*} 
& BugLocator & 0.27 & 0.49 & 0.60 & 0.31 & 0.37 \\
& DNNLOC & 0.46 & 0.71 & 0.78 & 0.41 & 0.51 \\
& KGBugLocator & - & - & - & - & - \\
& Dreamloc & 0.45 & 0.70 & 0.80 & 0.44 & 0.54 \\
& FLIM & 0.44 & 0.69 & 0.77 & 0.48 & 0.55 \\
& BL-GAN & 0.46 & 0.71 & 0.80 & 0.45 & - \\
& bjXnet & \textbf{0.60} & \textbf{0.83} & \textbf{0.90} & 0.61 & \textbf{0.70} \\
& MACL-IRFL & \textbf{0.60} & \textbf{0.83} & \textbf{0.90} & \textbf{0.64} & \textbf{0.70} \\
\midrule
\multirow{8}{*}{JDT}       
& BugLocator & 0.19 & 0.40 & 0.51 & 0.23 & 0.30 \\
& DNNLOC & 0.40 & 0.65 & 0.74 & 0.34 & 0.45 \\
& KGBugLocator & 0.44 & 0.69 & 0.76 & 0.40 & 0.48 \\
& Dreamloc & 0.50 & 0.77 & 0.87 & 0.52 & 0.64 \\
& FLIM & 0.38 & 0.65 & 0.76 & 0.41 & 0.50 \\
& BL-GAN & 0.42 & 0.66 & 0.74 & 0.39 & - \\
& bjXnet & 0.43 & 0.80 & 0.88 & 0.52 & 0.60 \\
& MACL-IRFL & \textbf{0.66} & \textbf{0.89} & \textbf{0.93} & \textbf{0.68} & \textbf{0.76} \\
\midrule
\multirow{8}{*}{SWT}       
& BugLocator & 0.19 & 0.38 & 0.51 & 0.25 & 0.28 \\
& DNNLOC & 0.35 & 0.69 & 0.80 & 0.37 & 0.45 \\
& KGBugLocator & 0.38 & 0.72 & 0.83 & 0.43 & 0.51 \\
& Dreamloc & 0.50 & 0.79 & 0.83 & 0.53 & 0.63 \\
& FLIM & 0.38 & 0.69 & 0.79 & 0.45 & 0.52 \\
& BL-GAN & - & - & - & - & - \\
& bjXnet & 0.52 & 0.79 & 0.88 & 0.55 & 0.65 \\
& MACL-IRFL & \textbf{0.65} & \textbf{0.87} & \textbf{0.90} & \textbf{0.67} & \textbf{0.75} \\
\midrule
\multirow{8}{*}{Tomcat}    
& BugLocator & 0.36 & 0.62 & 0.71 & 0.43 & 0.48 \\
& DNNLOC & 0.54 & 0.73 & 0.80 & 0.52 & 0.60 \\
& KGBugLocator & 0.51 & 0.75 & 0.83 & 0.56 & 0.62 \\
& Dreamloc & 0.53 & 0.77 & 0.85 & 0.55 & 0.62 \\
& FLIM & 0.50 & 0.73 & 0.81 & 0.53 & 0.60 \\
& BL-GAN & 0.53 & 0.73 & 0.82 & 0.56 & - \\
& bjXnet & - & - & - & - & - \\
& MACL-IRFL & \textbf{0.56} & \textbf{0.78} & \textbf{0.90} & \textbf{0.60} & \textbf{0.67} \\
\bottomrule
\end{tabular*}
\end{table}

\begin{figure*}
\centering

\begin{minipage}[t]{1\textwidth}
\centering
    \includegraphics[width=0.49\textwidth]{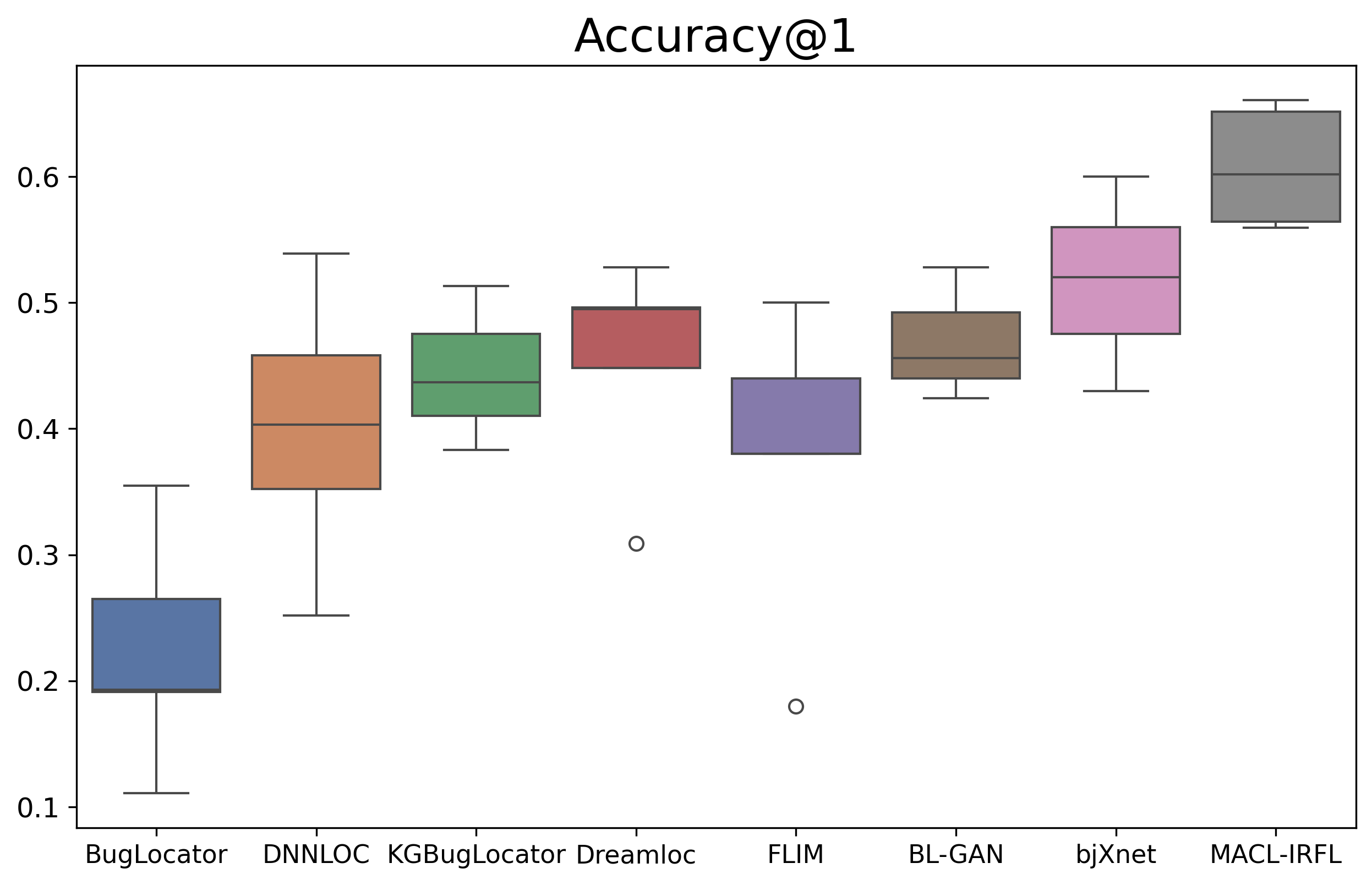}
    \includegraphics[width=0.49\textwidth]{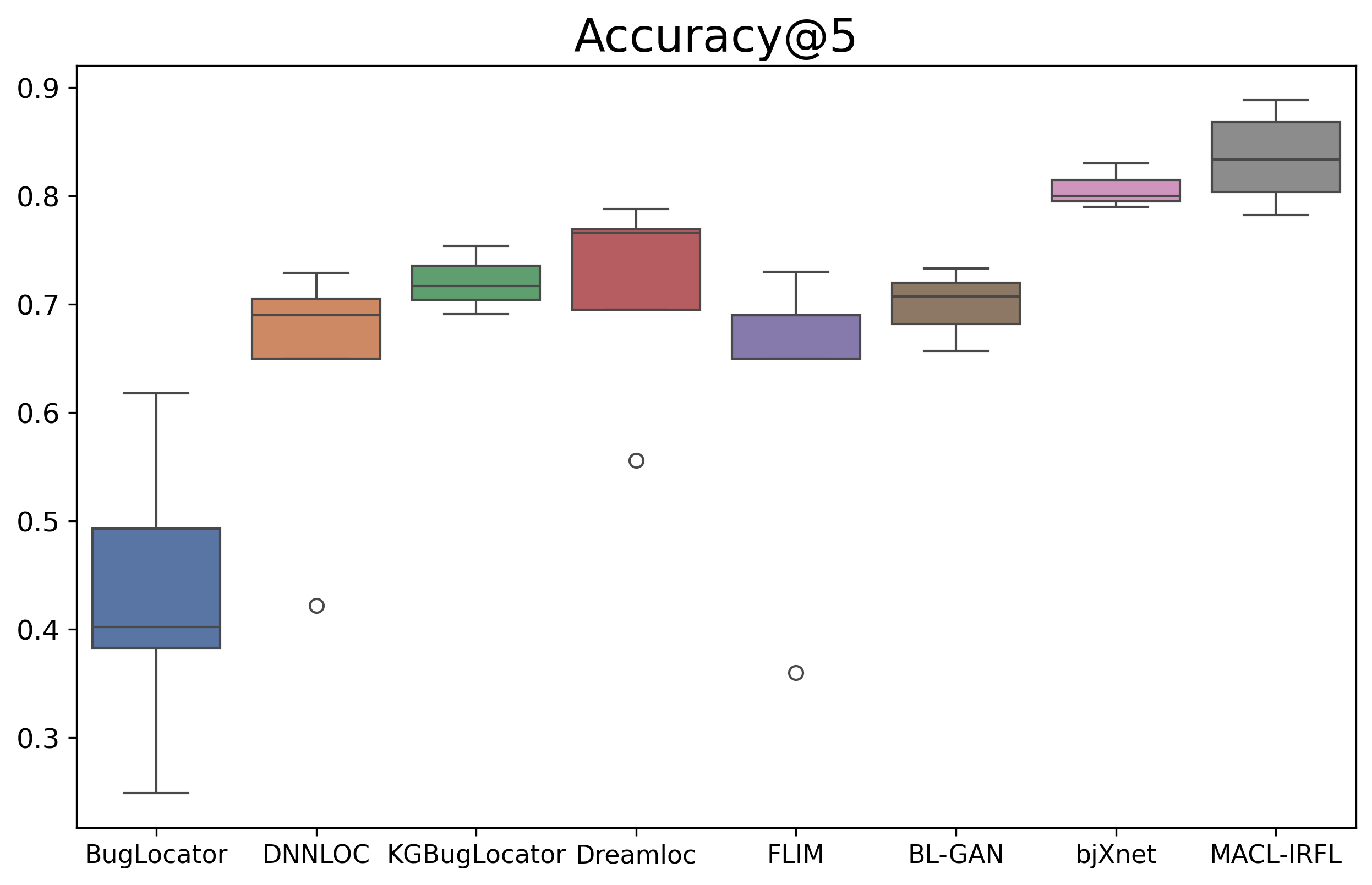}
\end{minipage}%

\begin{minipage}[t]{1\textwidth}
\centering
    \includegraphics[width=0.49\textwidth]{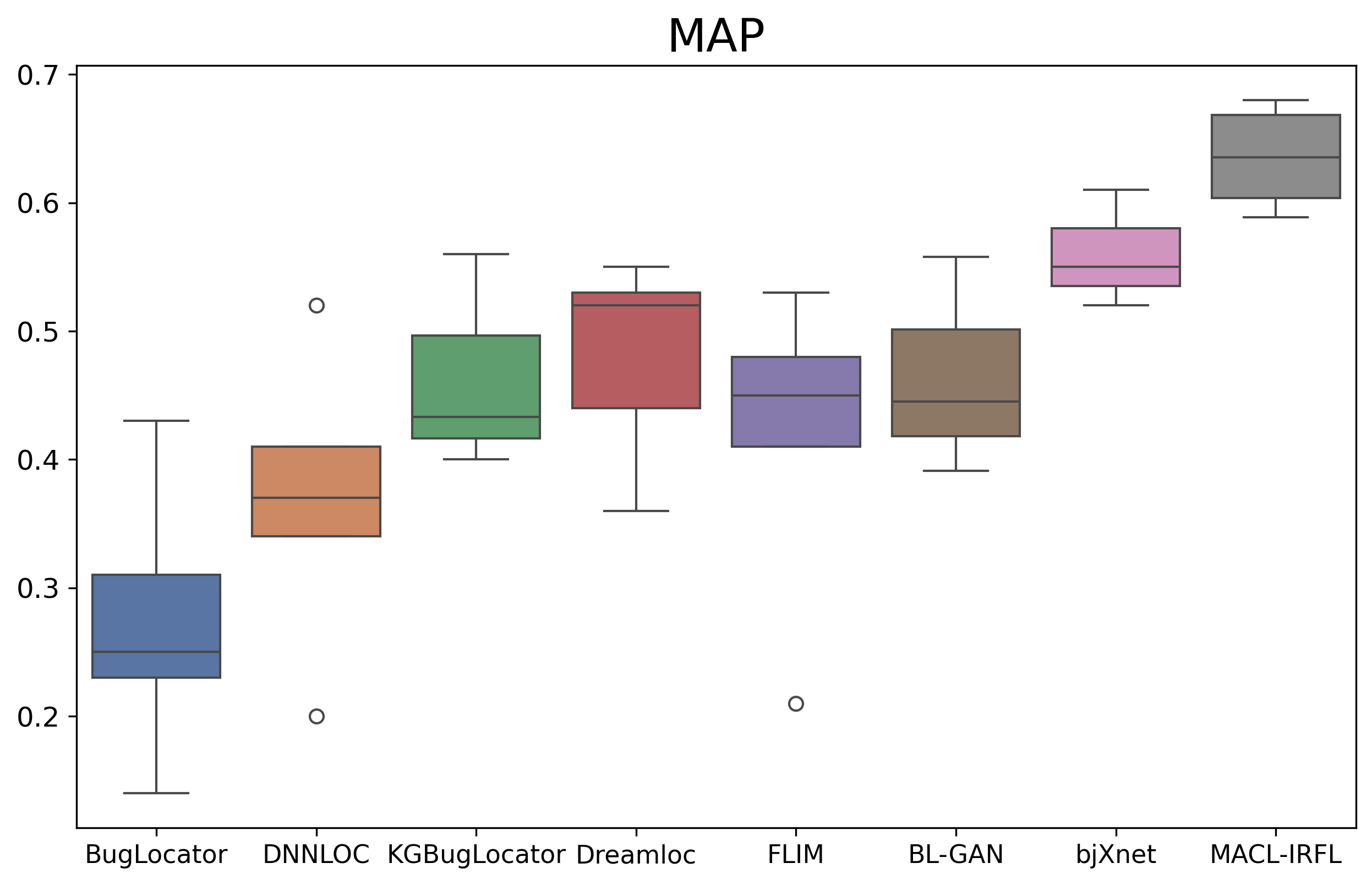}
    \includegraphics[width=0.49\textwidth]{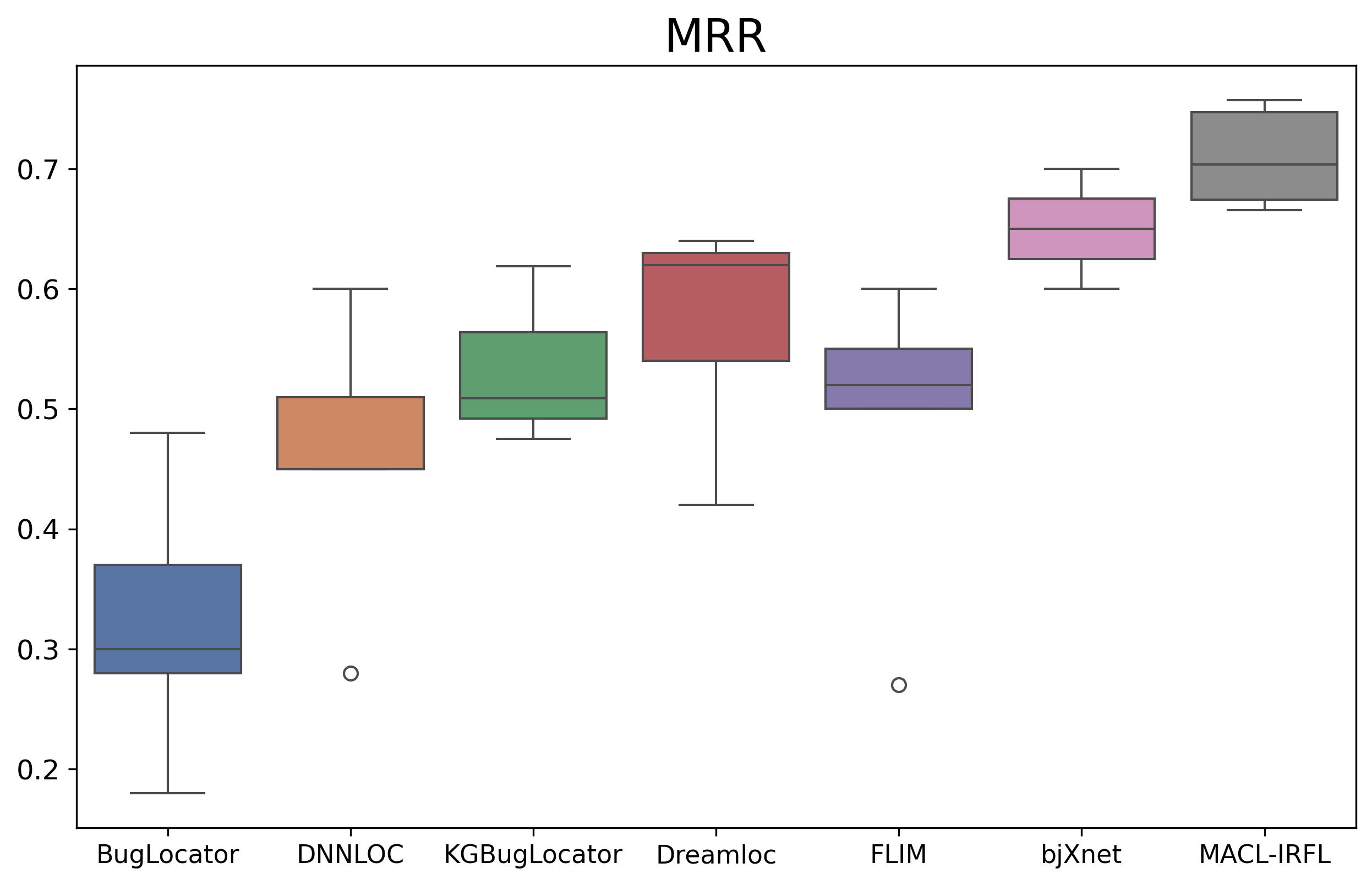}
\end{minipage}%

\caption{Performance comparison of eight approaches.}
\label{RQ1-performance-fig}
\end{figure*}

To evaluate the performance of our approach, we compare it to seven state-of-the-art fault localization approaches listed in Section \ref{baselines}, and calculate Accuracy@N, MAP and MRR, as shown in Table \ref{RQ1-performance-tab}. The best performance on each project is boldfaced. 
Fig. \ref{RQ1-performance-fig} shows a box-and-whisker plot illustrating the overall performance of the eight approaches.
Moreover, to verify the significant difference of the comparison results, we use Wilcoxon signed rank test and the effect size Cliff's $\delta$ to assess the performance difference between our approach and other baselines, as shown in Table \ref{RQ1-Wilcoxon-Cliff}.

For MAP and MRR score, our approach achieves 0.59-0.68 in MAP and 0.67-0.76 in MRR, which are the highest among all baselines. Compared with the corresponding best baseline model on the \textit{Birt}, \textit{Eclipse Platform UI}, \textit{JDT}, \textit{SWT} and \textit{Tomcat}, the MAP of our approach improves by 63.51\%, 4.19\%, 30.79\%, 21.55\% and 7.82\% respectively, and the MRR of our approach improves by 60.54\%, 0.55\%, 18.33\%, 14.95\% and 7.36\% respectively.
For Accuracy@N where $n$ is 1, 5, 10 in our experiments, we can find our approach achieves 0.56-0.66 in Accuracy@1, 0.78-0.89 in Accuracy@5, and 0.85-0.93 in Accuracy@10. 
Compared with the corresponding best baseline on each project, the Accuracy@1 of our approach improves by 80.97\% on the \textit{Birt} project, 0.27\% on \textit{Eclipse Platform UI} project, 33.47\% on \textit{JDT} project, 25.23\% on \textit{SWT} project and 4.70\% on \textit{Tomcat} project, respectively.
In other words, when considering only the first ranked candidate in the results returned by the model, our model is able to provide better localization results.
According to the statistical test results shown in Table \ref{RQ1-Wilcoxon-Cliff}, the results demonstrate that, with the exception of bjXnet, our approach performs significantly better than other baselines in most cases at the confidence level of 95\% (i.e., $p$-value\textless 0.05) and $|\delta| \ge 0.474$. Although there is no significant difference between our approach and bjXnet, our results are still competitive in Accuracy@1, MAP and MRR in terms of Cliff's $\delta$ value ($|\delta| \ge 0.474$). More specifically, according to the experimental results, we have the following observations.

\begin{sidewaystable}
\caption{Statistical test results of our approach vs. other approaches based on Wilcoxon signed-rank test and Cliff delta, where avg denotes the average result of each approach on five projects. 
L and M denote the large (i.e., $|\delta| \ge 0.474$) and medium (i.e., $0.33 \leq |\delta| < 0.474$) difference between our model and other approaches.
}
\label{RQ1-Wilcoxon-Cliff}
\begin{tabular*}{\textwidth}{@{\extracolsep\fill}ccccccccccc}
\toprule
\multicolumn{1}{c}{Our approach} & \multicolumn{2}{c}{Accuracy@1} & \multicolumn{2}{c}{Accuracy@5} & \multicolumn{2}{c}{Accuracy@10} & \multicolumn{2}{c}{MAP} & \multicolumn{2}{c}{MRR} \\ 
\textit{vs} & avg & p-value ($\delta$) & avg & p-value ($\delta$) & avg & p-value ($\delta$) & avg & p-value ($\delta$) & avg & p-value ($\delta$) \\
\midrule
BugLocator & 0.223 & $\textbf{p\textless 0.05}$ (L) & 0.429 & $\textbf{p\textless 0.05}$ (L) & 0.531 & $\textbf{p\textless 0.05}$ (L) & 0.272 & $\textbf{p\textless 0.05}$ (L) & 0.322 & $\textbf{p\textless 0.05}$ (L) \\
DNNLOC & 0.401 & $\textbf{p\textless 0.05}$ (L) & 0.639 & $\textbf{p\textless 0.05}$ (L) & 0.728 & $\textbf{p\textless 0.05}$ (L) & 0.368 & $\textbf{p\textless 0.05}$ (L) & 0.458 & $\textbf{p\textless 0.05}$ (L) \\
KGBugLocator & 0.444 & $\textbf{p\textless 0.05}$ (L) & 0.721 & $\textbf{p\textless 0.05}$ (M) & 0.807 & $\textbf{p\textless 0.05}$ (L) & 0.464 & $\textbf{p\textless 0.05}$ (L) & 0.534 & $\textbf{p\textless 0.05}$ (L)\\
Dreamloc & 0.455 & $\textbf{p\textless 0.05}$ (L) & 0.715 & $\textbf{p\textless 0.05}$ (L) & 0.800 & $\textbf{p\textless 0.05}$ (L) & 0.480 & $\textbf{p\textless 0.05}$ (L) & 0.570 & $\textbf{p\textless 0.05}$ (L)\\
FLIM & 0.376 & $\textbf{p\textless 0.05}$ (L) & 0.624 & $\textbf{p\textless 0.05}$ (L) & 0.716 & $\textbf{p\textless 0.05}$ (L) & 0.416 & $\textbf{p\textless 0.05}$ (L) & 0.488 & $\textbf{p\textless 0.05}$ (L) \\
BL-GAN & 0.469 & $\textbf{p\textless 0.05}$ (L) & 0.699 & $\textbf{p\textless 0.05}$ (L) & 0.788 & $\textbf{p\textless 0.05}$ (L) & 0.465 & $\textbf{p\textless 0.05}$ (L) & - & - \\
bjXnet & 0.517 & p\textgreater 0.05 (L) & 0.807 & p\textgreater 0.05 (M) & 0.887 & p\textgreater 0.05 (M) & 0.560 & p\textgreater 0.05 (L) & 0.650 & p\textgreater 0.05 (L) \\
MACL-IRFL & 0.607 & - & 0.835 & - & 0.901 & - & 0.635 & - & 0.710 & - \\
\bottomrule
\end{tabular*}
\end{sidewaystable}

We found that among all compared approaches, BugLocator does not achieve good performance in most cases.
Compared with it, our approach shows distinct improvements in terms of all metrics on all projects. This may be because that our approach takes advantage of historical bug-fix records to predict the potential relevancy between bug reports and source code files, rather than simply calculating the similarity between an individual pair of bug report and code file, which can easily be affected by the quality of report text.
For deep learning based approaches, compared with the DNNLOC, our approach achieves around 0.27 and 0.25 improvement in MAP and MRR on average of the five datasets. 
DNNLOC combines three types of features (relevancy computed from DNN, textual similarity computed via rVSM, and the metadata of the bug-fixing history) to fully compute the relevancy between bug reports and source code files. 
Dreamloc, FLIM, KGBugLocator, and bjXnet mainly focus on enriching semantic information to bridge the lexical gap. Dreamloc and FLIM split source code files into code snippets and functions, respectively. KGBuglocator constructs code knowledge graph based on abstract syntax tree (AST) of every source files and utilizes keywords supervised constraint to enhance model with interactive information between source files and bug reports. bjXnet builds CPGs (consists of AST, CFG, and data dependence graph (DDG)) of source code, which can well reflect the information on the source code structure, the statement execution process, the control dependence, and the data dependence, so as to supplement the semantic information of the source code.
The average scores of these four approaches on the five datasets in MAP and MRR are better than DNNLOC, which indicates that the semantic information is very important to understand programming language.
Although extracting richer semantic information or adding additional features can bridge the gap between natural language text and code language to some extent, these deep learning approaches learn the features of bug reports and source code files separately.
In contrast to these baselines, our proposed approach MACL-IRFL leverages a multi-view learning framework to capture the interactive information between bug reports and source code files. Consequently, the features learned from bug reports inherently encompass information from the source code files, enhancing the relevance during the matching process. 
Moreover, the bug information is often located in a code block, so the entire code file may have a lot of non-bug-related information. 
Although KGBugLocator pays attention to the interaction information between bug reports and source code files, the interaction information via keyword matching is still affected by the quality of the text. Dreamloc and FLIM, on the other hand, split the source code into finer granularity to match with report, they certainly increase the computational cost of matching.
In the three views we construct, the initial features of bug reports and source code files are randomly initialized, which are not sensitive to text quality of the reports or the programming code of the source files.
Our approach is to aggregate all buggy files information into their corresponding bug report through the interactions in the network, which can make the learned features of bug reports are highly correlated to their buggy files (i.e., ground truth) when training.
Hence, MACL-IRFL with its interactive feature learning, achieves more robust and comprehensive matching and is less sensitive to the intrinsic limitations of bug report quality.
Compared with the BL-GAN, MACL-IRFL achieves around 0.18 improvement in MAP on average of the three datasets (i.e., \textit{Eclipse Platform UI}, \textit{JDT}, \textit{Tomcat}). 
BL-GAN employs GAN to generate report-code pairs in the case of insufficient bug-fix records. 
MACL-IRFL addresses the case of insufficient historical bug-fix records by adding additional relationships, i.e., similarity relationships between bug reports and co-citation relationships between source code files. Meanwhile, MACL-IRFL alleviates report-code interaction sparsity, and avoids the noise that might be caused by fake report-code pairs.

Among all compared baselines, our approach achieves the best performance in most cases across all those projects. The main reasons are three folds: 
1) \textbf{MACL-IRFL constructs a multi-view framework that explicitly encodes the intricate interaction features between bug reports and source code files into their respective representation learning processes}, which achieves a deep integration of information and gives higher ranking scores.
2) \textbf{MACL-IRFL does not pay much attention on the initial natural language text in bug reports and programming codes in source files} since we assign random initial features to the bug reports and the source code files, which could mitigate the problem of low-quality bug reports and noisy programming codes.
3) \textbf{MACL-IRFL incorporates and encodes the similarity relationships and the co-citation relationships into the representation learning of bug reports and source code files with weights}, which helps to alleviate the insufficient bug-fix records issue and improve the accuracy of localization.

\textbf{RQ2: How do adaptive contrastive learning strategy impact our model?}

\begin{figure*}
\centering
\begin{minipage}[t]{1\textwidth}
\centering
    \includegraphics[width=0.19\textwidth]{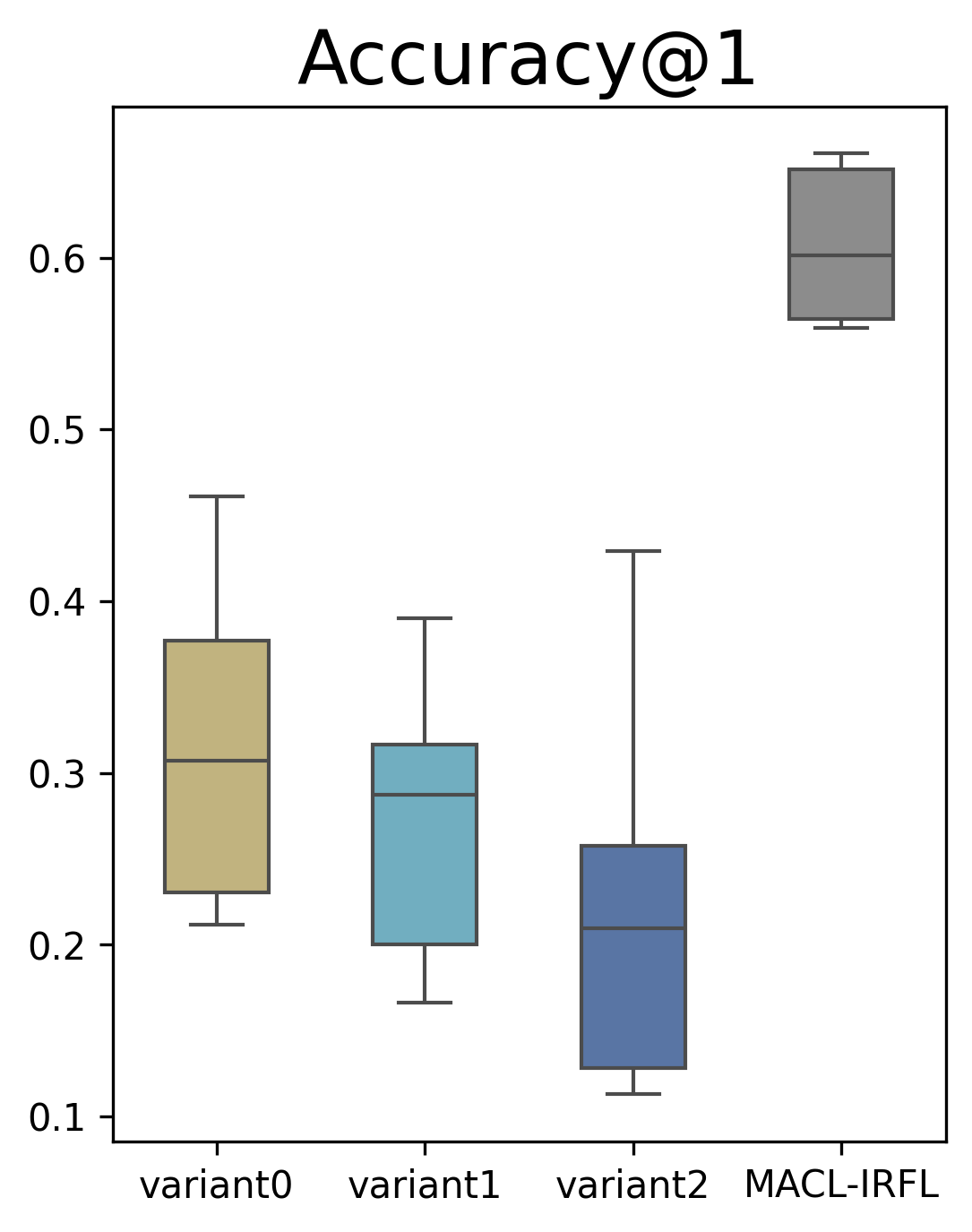}
    \includegraphics[width=0.19\textwidth]{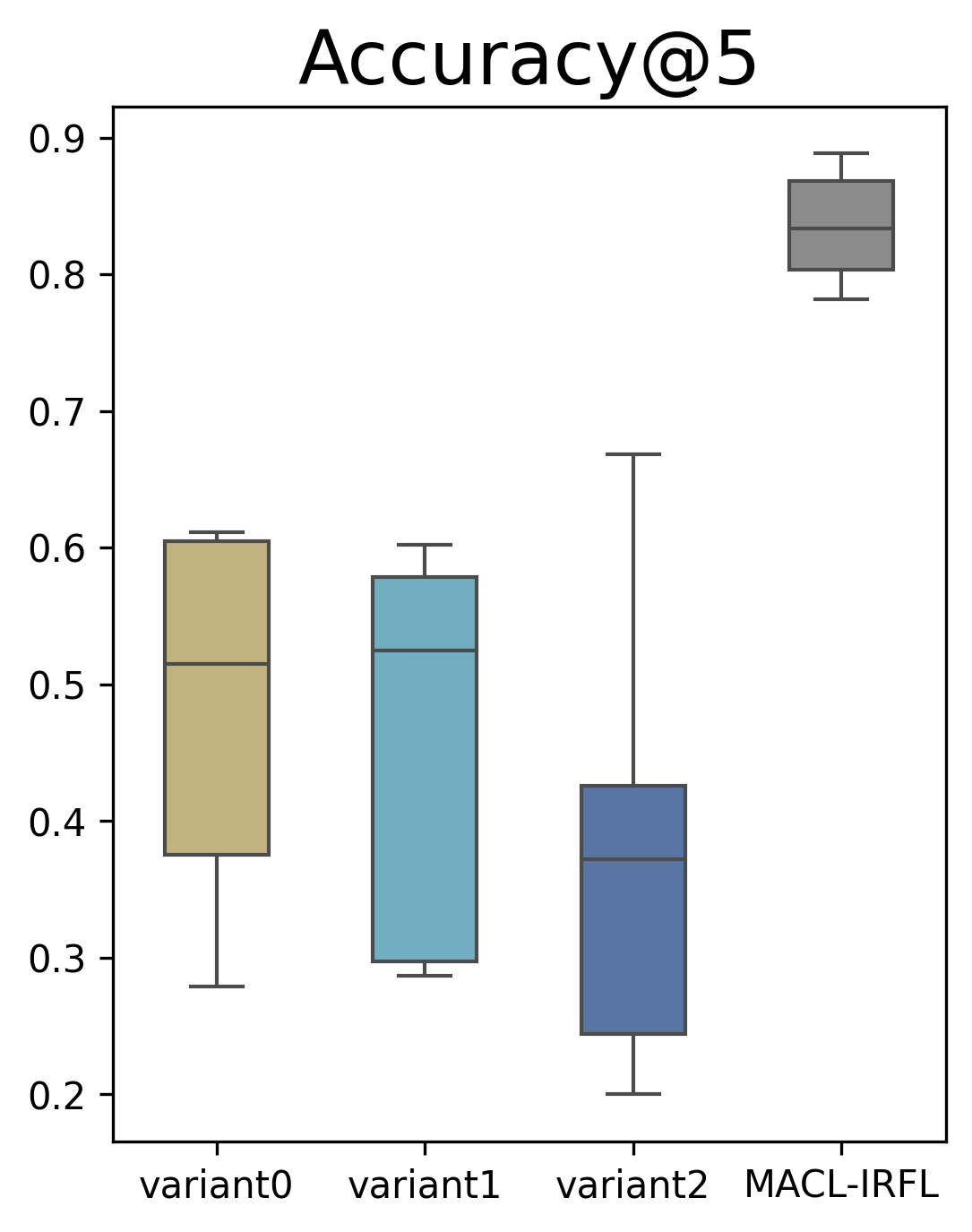}
    \includegraphics[width=0.19\textwidth]{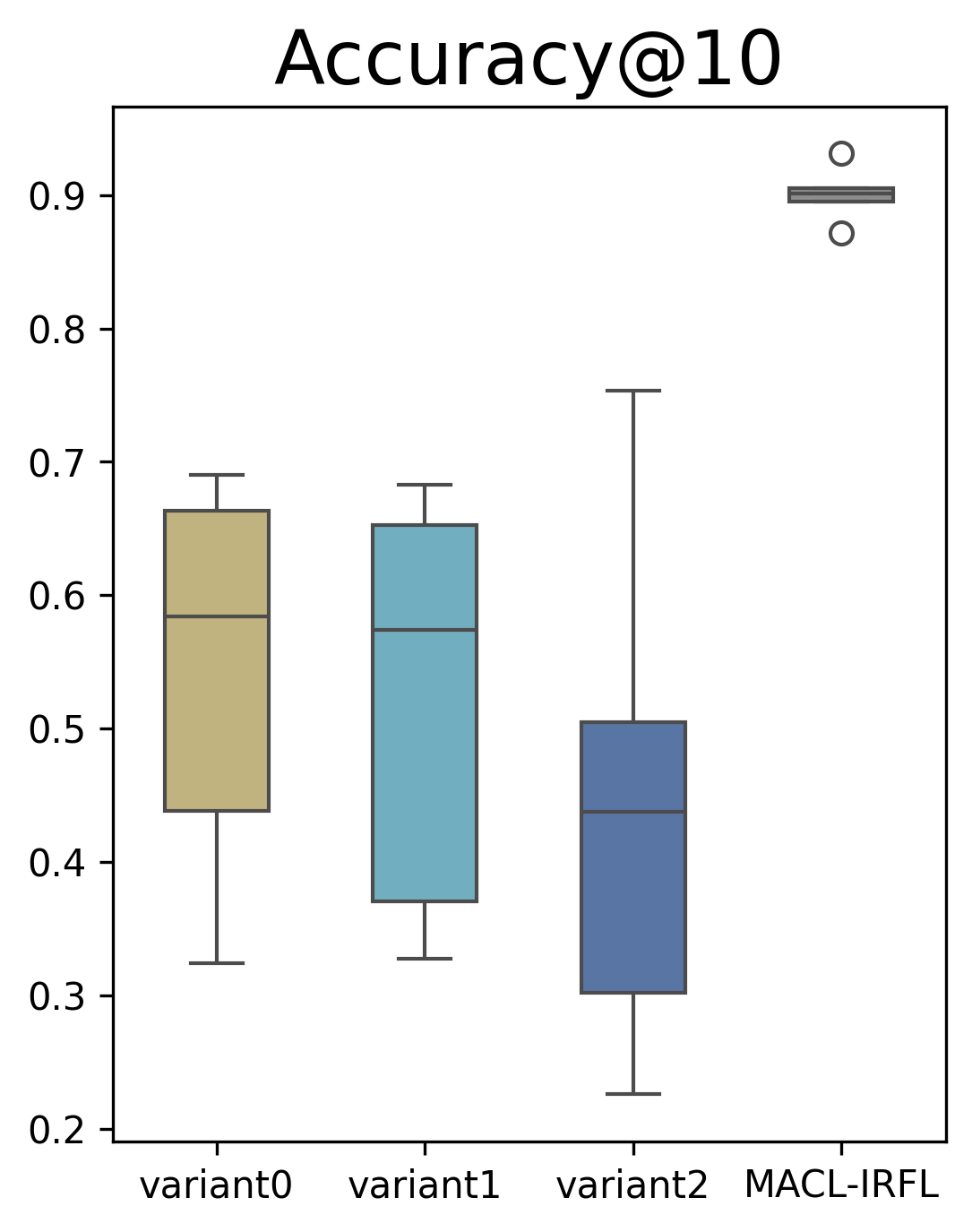}
    \includegraphics[width=0.19\textwidth]{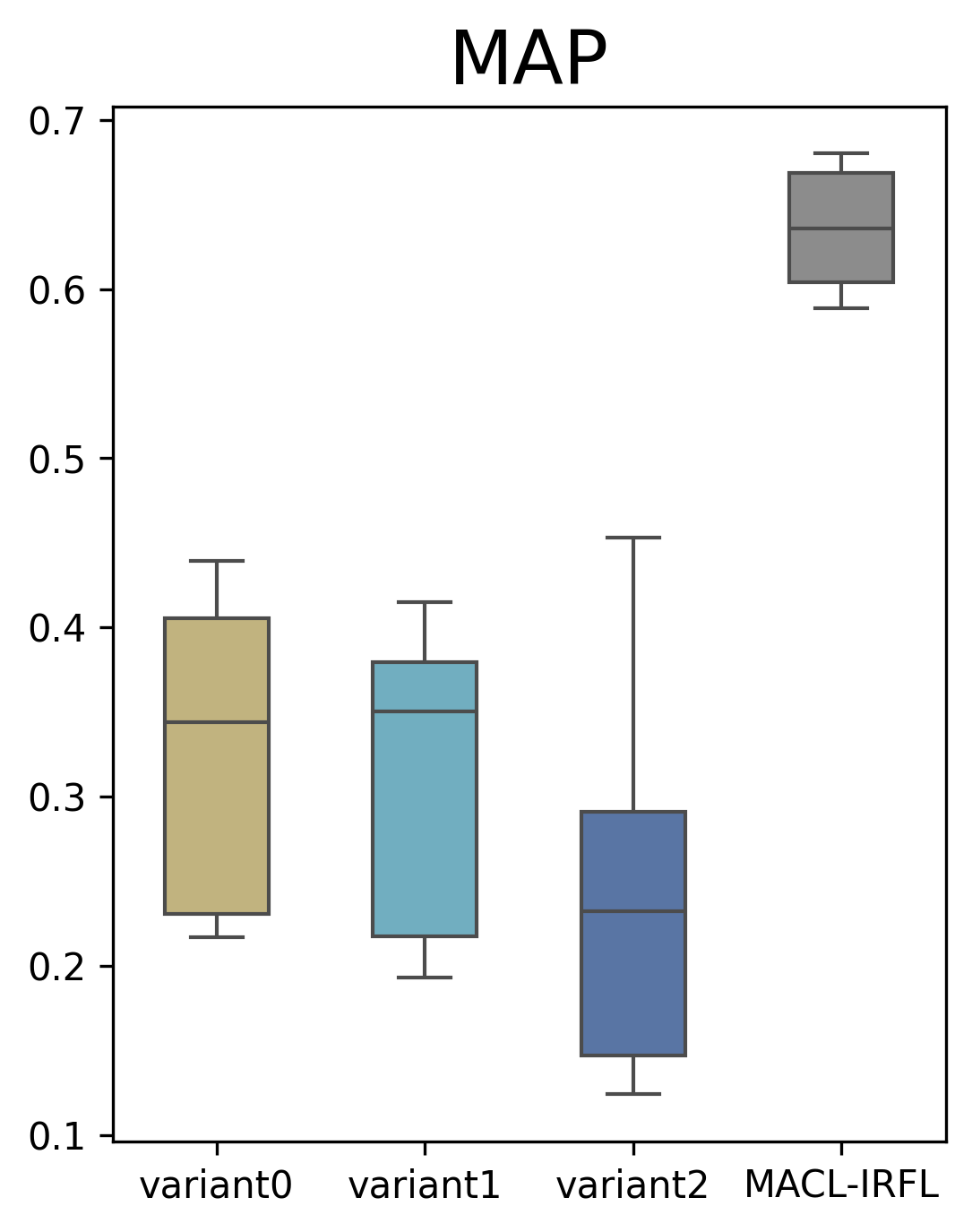}
    \includegraphics[width=0.19\textwidth]{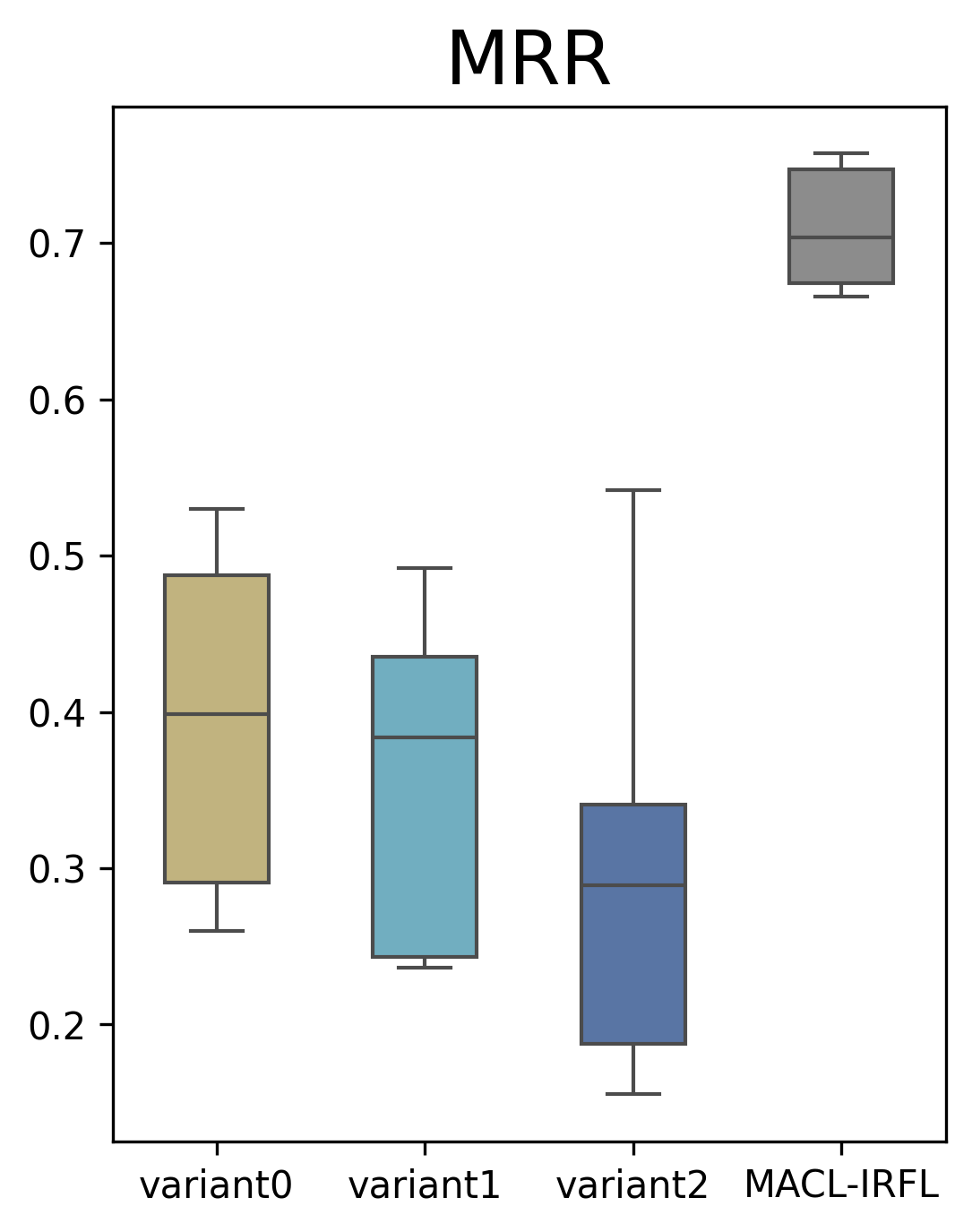}
\end{minipage}%
\caption{Comparison of the impacts of different variants on performance.}
\label{RQ2-performance}
\end{figure*}

\begin{figure*}
\centering
\begin{minipage}[t]{1\textwidth}
\centering
    \includegraphics[width=0.24\textwidth]{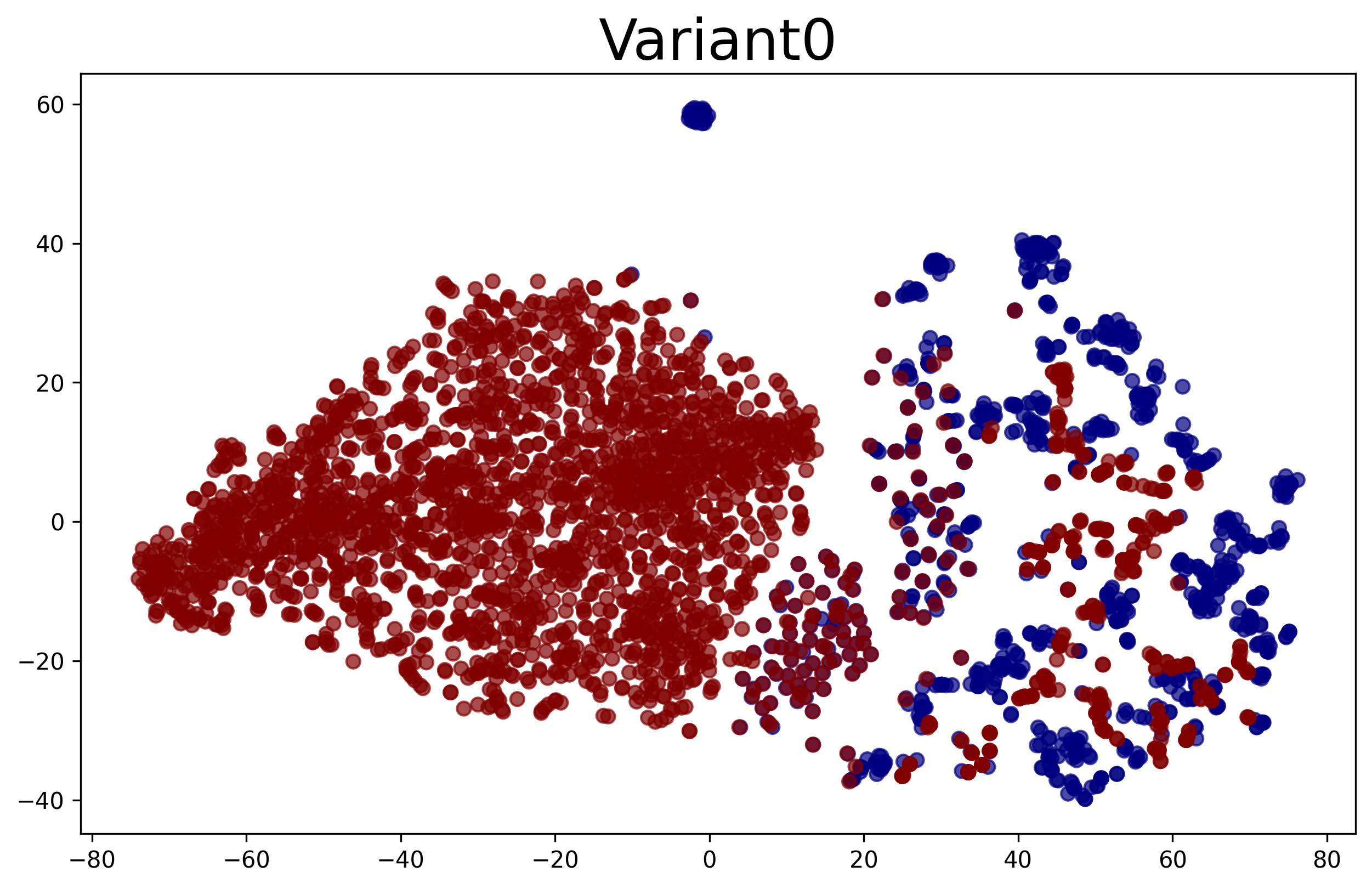}
    \includegraphics[width=0.24\textwidth]{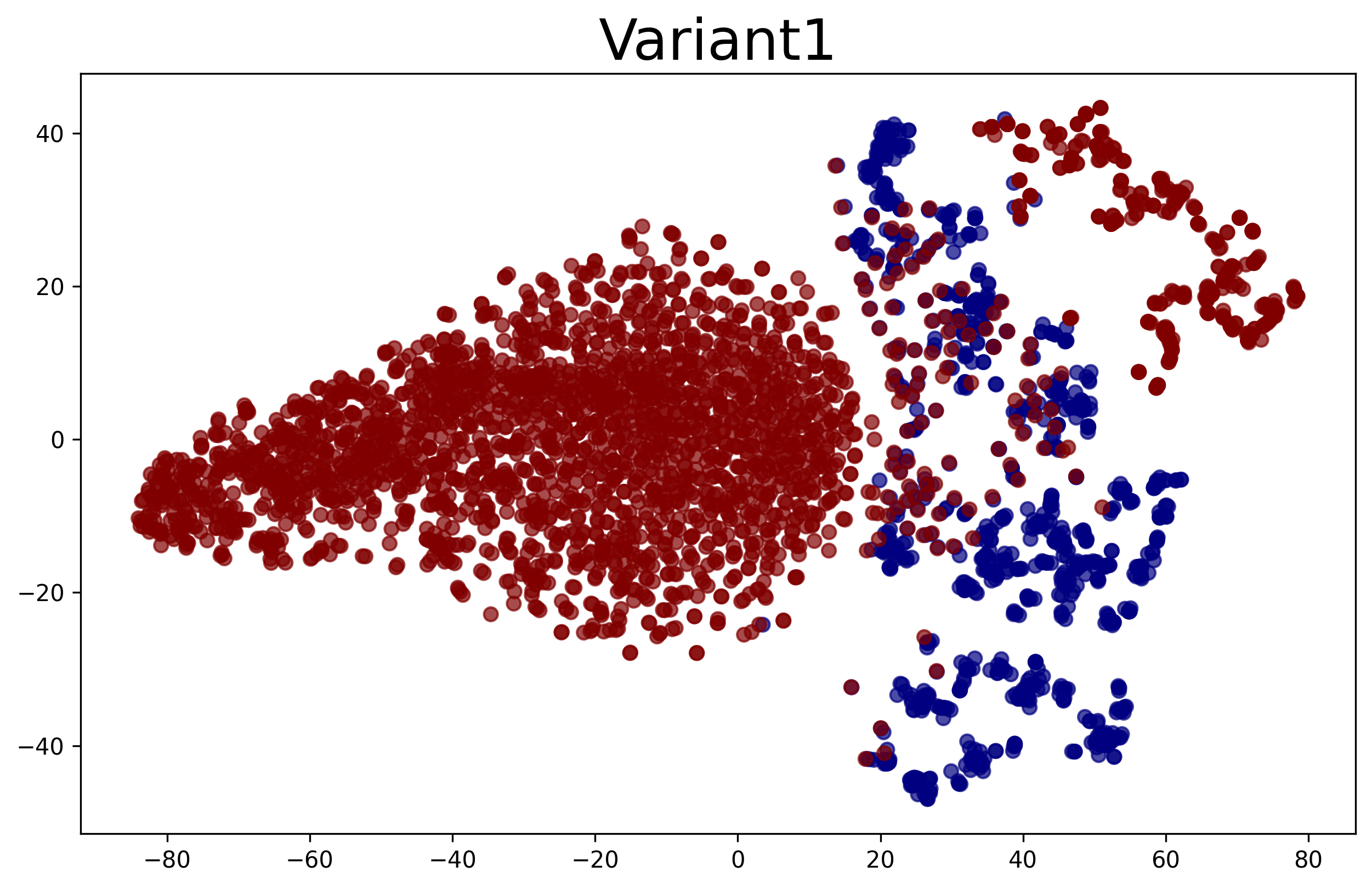}
    \includegraphics[width=0.24\textwidth]{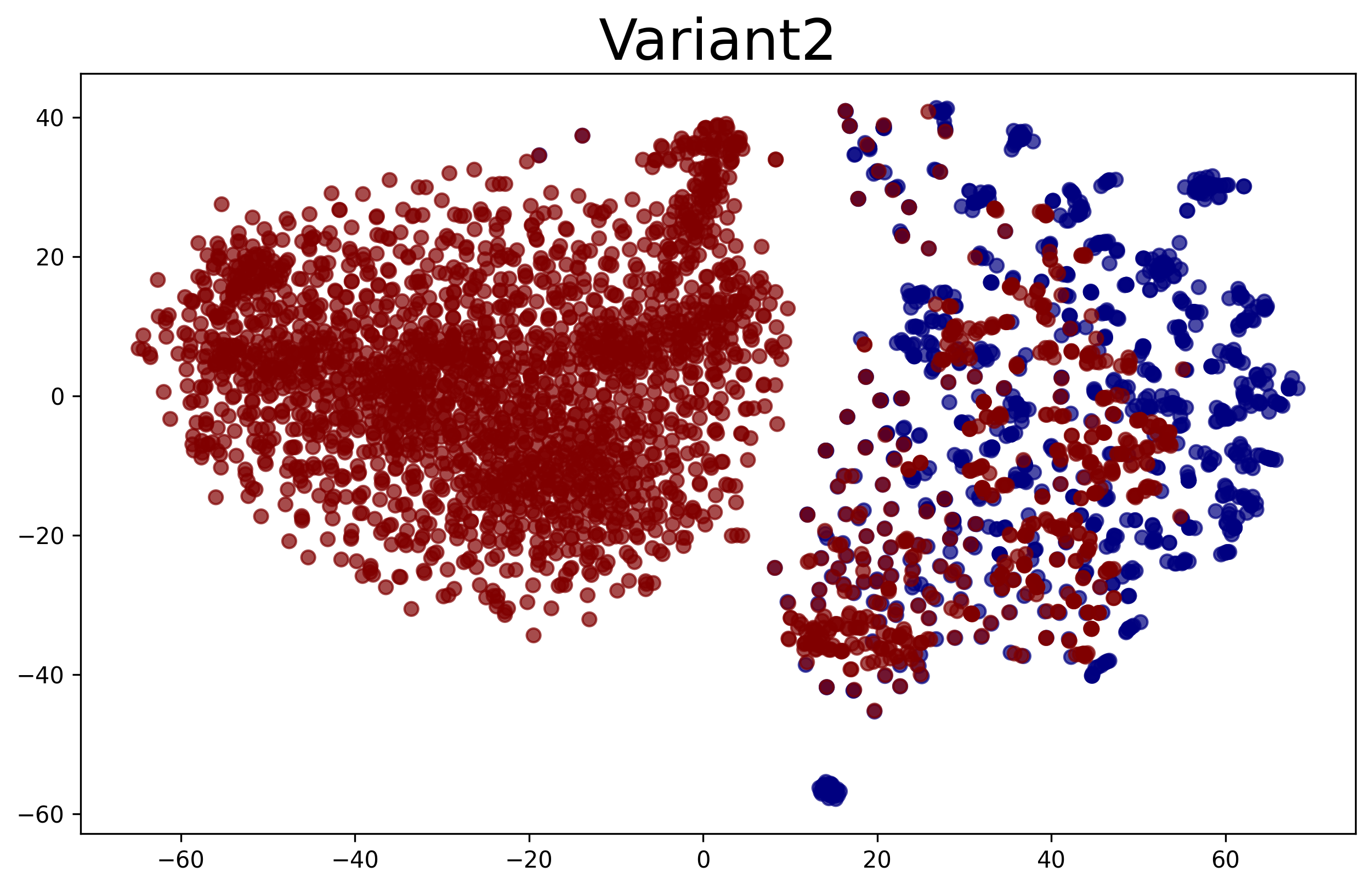}
    \includegraphics[width=0.24\textwidth]{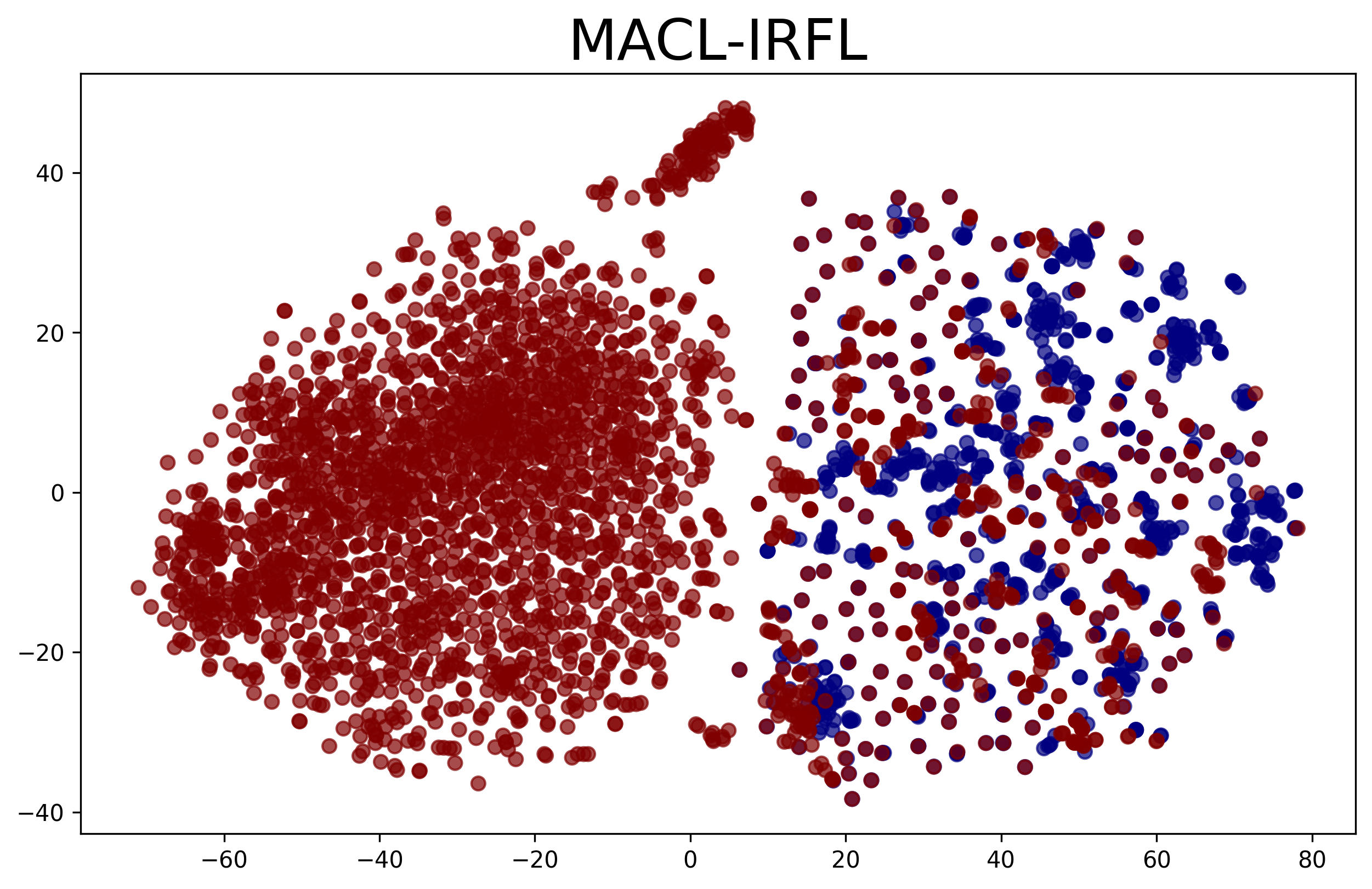}
\end{minipage}%
\caption{Two dimensional projections of t-SNE \cite{van2008visualizing} visualization embeddings of different variants for \textit{Tomcat} dataset. The red nodes and bule nodes denote the distribution of source code files and bug reports respectively.}
\label{RQ2-feature-distribution}
\end{figure*}

In this experiment, we complete a comprehensive ablation study to investigate the impact of integrating the contrastive learning strategy on the localization performance of our model. Here, we construct three variants of MACL-IRFL as shown in Table \ref{variants}. 
Fig. \ref{RQ2-performance} shows the comparison results, and the statistical test results also show that MACL-IRFL performs significantly better than other variants ($p$-value\textless 0.05 and $|\delta| \geq$ 0.474).
We take \textit{Tomcat} as an example to visualize the three variants and MACL-IRFL representation space, as shown in Fig. \ref{RQ2-feature-distribution}. 
We have the following observations.

First, we found that \textbf{$variant_0$ achieves poor performance than MACL-IRFL.} The average Accuracy@1, Accuracy@5, Accuracy@10, MAP, and MRR of $variant_0$ are 31.74\%, 47.71\%, 54.00\%, 32.73\% and 39.33\%, which are 29.00\%, 35.82\%, 36.07\%, 30.80\% and 31.63\% lower than MACL-IRFL, respectively. This demonstrates that performing multi-view framework with contrastive learning task achieves better performance than fault localization task alone. There are two main reasons for the inferior performance of $variant_0$.
On the one hand, the node representations in the validation and test sets lack the bug reports and source code files interaction information. As we mentioned above, the history bug-fixing records are only accessible during the training phase. Thus, even though the bug report representations can learn interaction information with source code files during training, it still significantly constrained for the bug report nodes encountered in the validation and test sets.
On the other hand, given that the training set exhibits a bipartite graph structure of report-code interaction relationships, whereas the validation and test sets are comprised of homogeneous graphs of report-report and code-code views, respectively, there exists a considerable distribution gap. Consequently, the parameters learned from the training set often fail to transfer effectively to the distributions of the validation and test sets.

Second, $variant_1$ incorporates the similarity relationships (i.e., report-report view), while $variant_2$ incorporates the co-citation relationships (i.e., code-code view) on the basis of $variant_0$ in training phase, respectively. The results show that \textbf{$variant_1$ and $variant_2$ achieve inferior performance than $variant_0$ in most cases.} The average Accuracy@1, Accuracy@5, Accuracy@10, MAP, and MRR of $variant_1$ are 27.20\%, 45.80\%, 52.15\%, 31.09\% and 35.82\%, which are 4.54\%, 1.91\%, 1.86\%, 1.64\% and 3.51\% lower than $variant_0$. 
While the average Accuracy@1, Accuracy@5, Accuracy@10, MAP, and MRR of $variant_2$ are 22.76\%, 38.21\%, 44.49\%, 24.94\% and 30.30\%, which are 8.99\%, 9.50\%, 9.52\%, 7.79\% and 9.04\% lower than $variant_0$. 
The results demonstrate that conducting contrastive learning solely on a single node type (either reports or code) yields suboptimal performance, failing to enhance overall performance.  
We attribute this deterioration to the neglect of the intricate interdependence and complementary between node types, inadvertently amplifying feature distance disparities between bug report and source code nodes. 
\textbf{$variant_1$ and $variant_2$, in an attempt to distinguish between different views or nodes, may overemphasize intra-type differences while overlooking the cross-type similarities and connections.} This narrowed perspective not only fails to bridge the gap between report and code features but potentially widens it, adversely affecting performance on tasks that require an integrated understanding of both.

Third, in addition to the quantitative results above, we present an alternative qualitative analysis by directly visualizing the representation space and seeing the layout of the cluster for the three variants and MACL-IRFL. From the Fig. \ref{RQ2-feature-distribution}, the red nodes denote source code files, and blue nodes denote bug reports. Notably, those bug-fixing source code files that interact with bug reports overlap with the bug report cluster, while those clean files (irrelevant negative samples) are distant to the bug report cluster. 
\textbf{An observation across $variant_0$, $variant_1$ and $variant_2$ reveals that even within the overlapping regions, the distance between bug reports and source code nodes remains relatively dispersed. In contrast, MACL-IRFL exhibits a more uniform distribution, effectively narrowing the feature distance between bug reports and source code files.} Therefore, the adaptive data augmentation is necessary in fault localization scenario, which can effectively drop out task-irrelevant information and alleviate auxiliary information overload problem.

\textbf{RQ3: Does the embedding propagation from GNN high-layer help improve the localization performance?}

\begin{table}[h]
\caption{Effect of aggregation layer numbers ($L$).}
  \label{RQ3-average}
\begin{tabular*}{\textwidth}{@{\extracolsep\fill}cccccc}
\toprule
\multirow{2}{*}{Layer numbers} & \multicolumn{3}{c}{Accuracy@N} & \multirow{2}{*}{MAP} & \multirow{2}{*}{MRR} \\\cmidrule{2-4} 
& n=1 & n=5 & n=10 & & \\
\midrule
MACL-IRFL-1 & 0.174 & 0.329 & 0.388 & 0.213 & 0.250 \\
MACL-IRFL-2 & 0.412 & 0.613 & 0.692 & 0.442 & 0.507 \\
MACL-IRFL-3 & 0.573 & 0.818 & 0.878 & 0.607 & 0.682 \\
MACL-IRFL-4 & 0.546 & 0.777 & 0.856 & 0.582 & 0.654 \\
MACL-IRFL-5 & 0.473 & 0.748 & 0.825 & 0.528 & 0.596 \\
\bottomrule
\end{tabular*}
\end{table}

\begin{figure*}
\centering

\begin{minipage}[t]{1\textwidth}
\centering
    \includegraphics[width=0.49\textwidth]{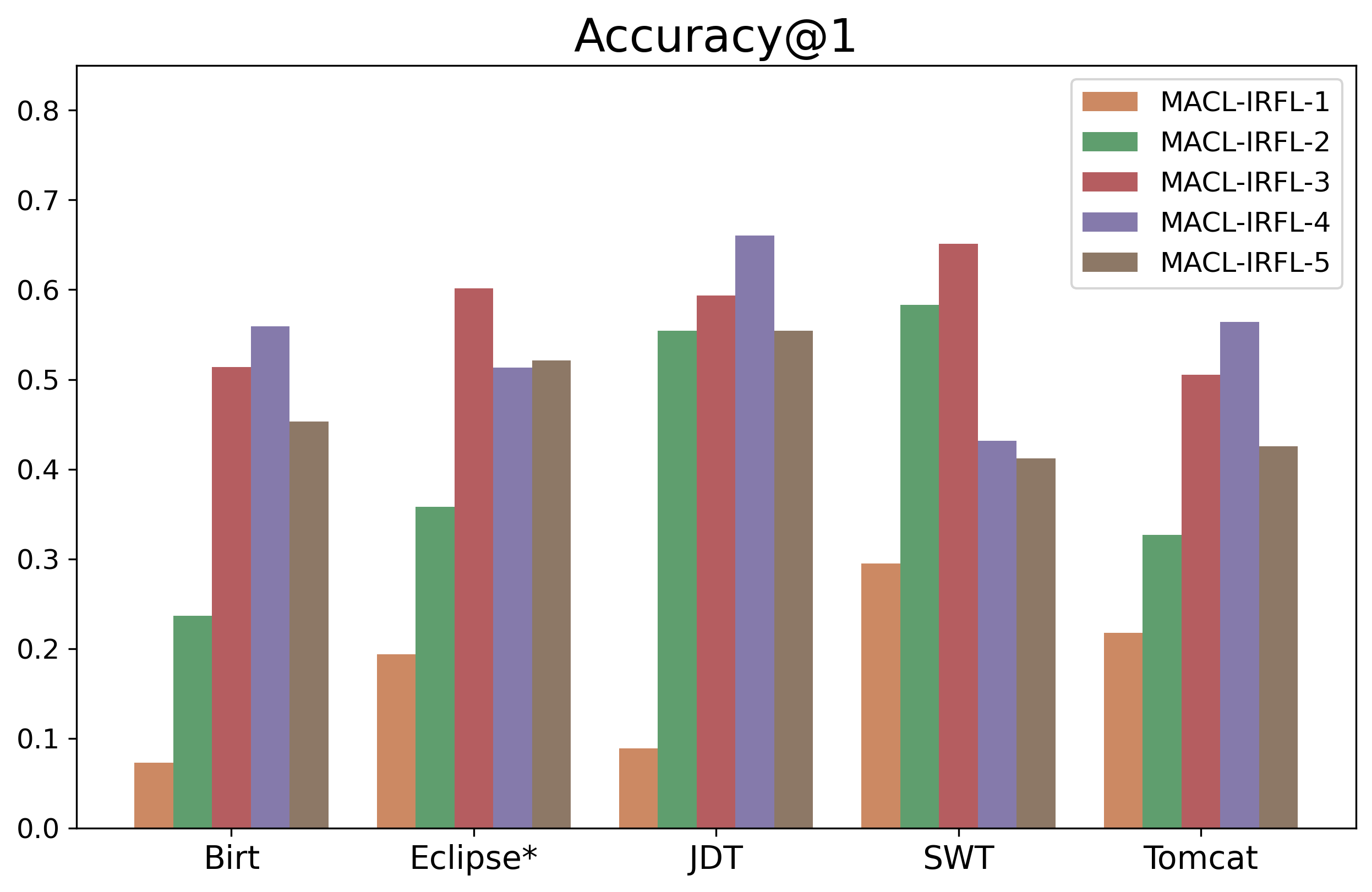}
    \includegraphics[width=0.49\textwidth]{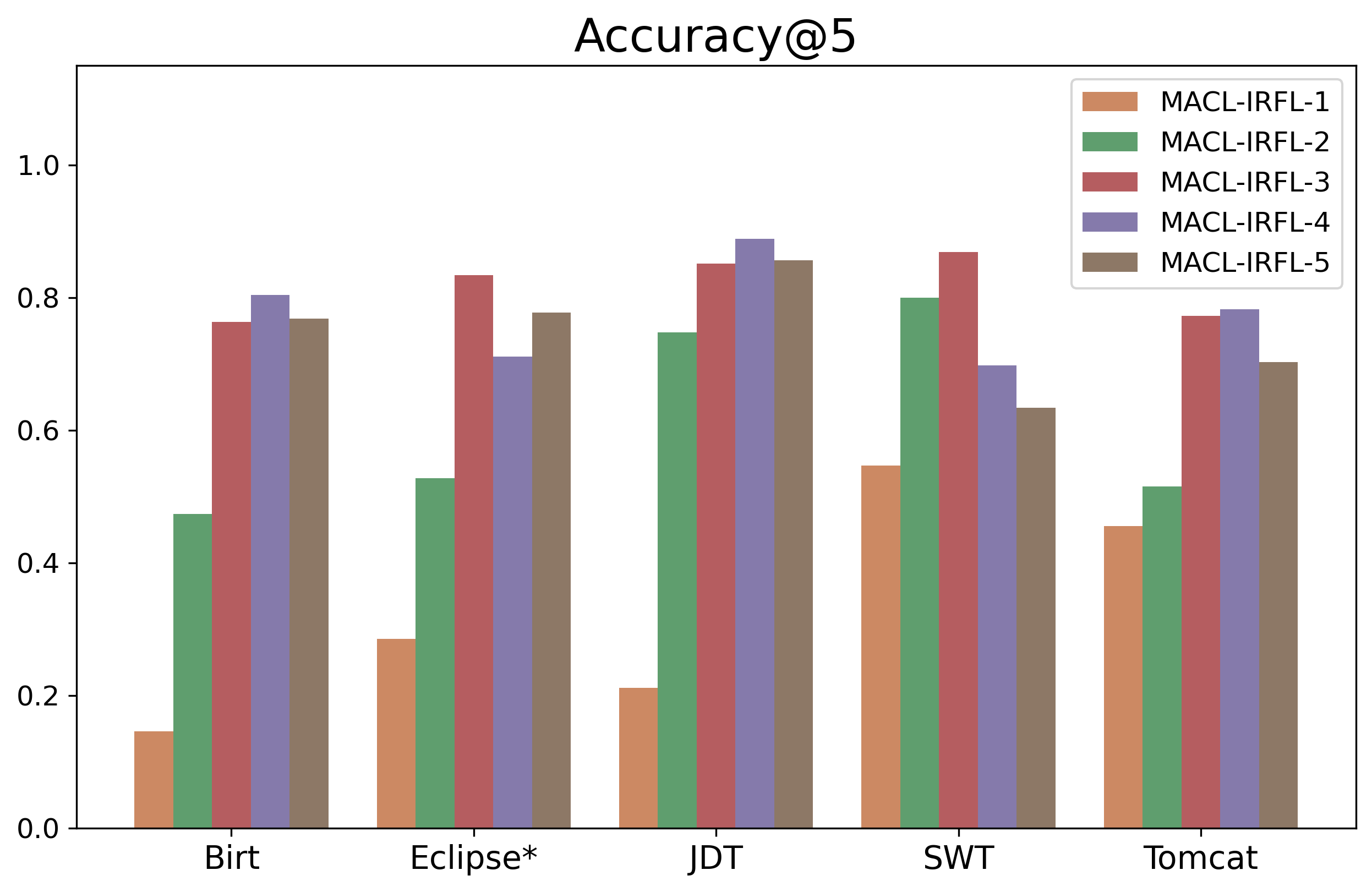}
\end{minipage}%

\begin{minipage}[t]{1\textwidth}
\centering
    \includegraphics[width=0.49\textwidth]{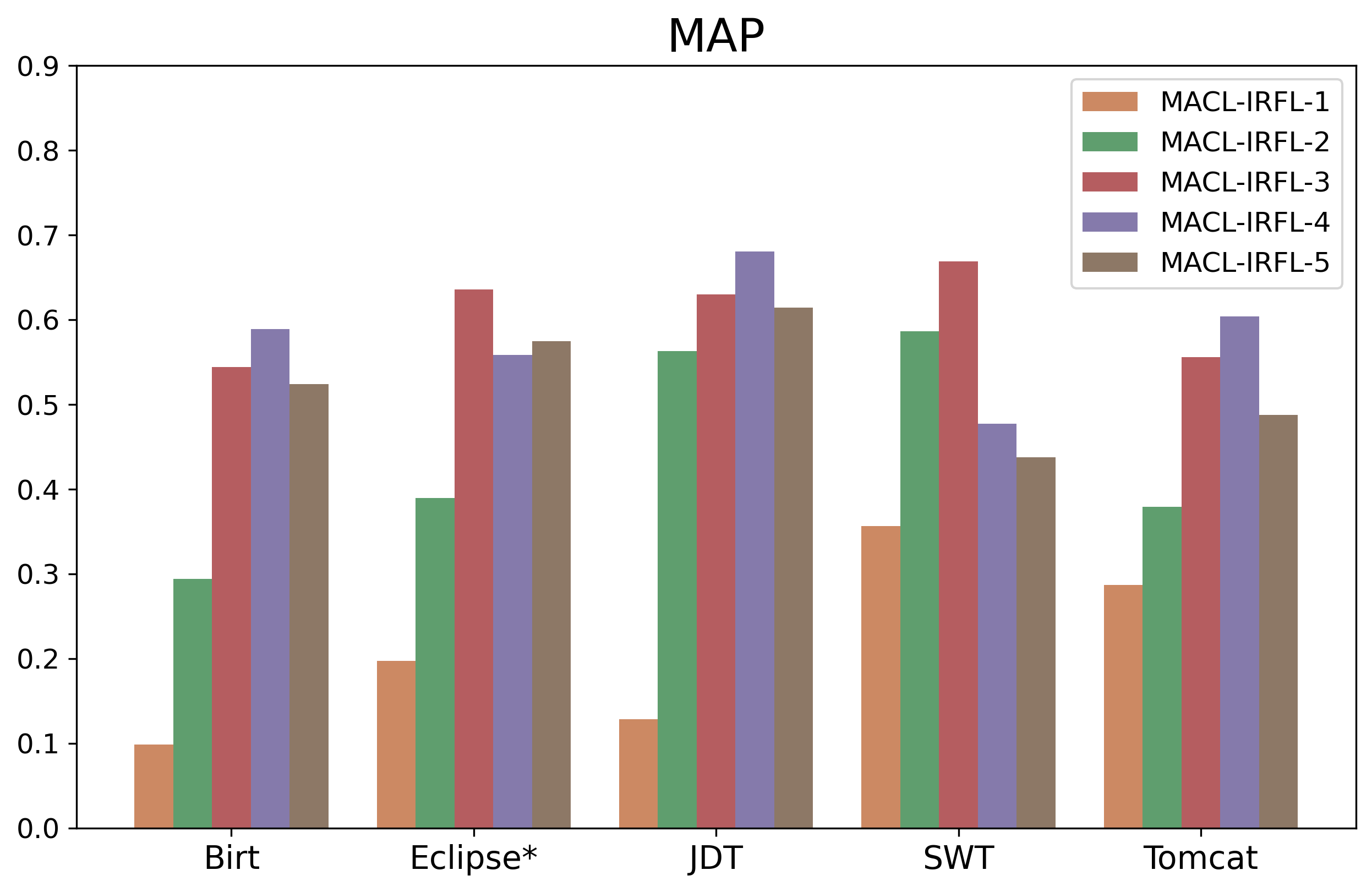}
    \includegraphics[width=0.49\textwidth]{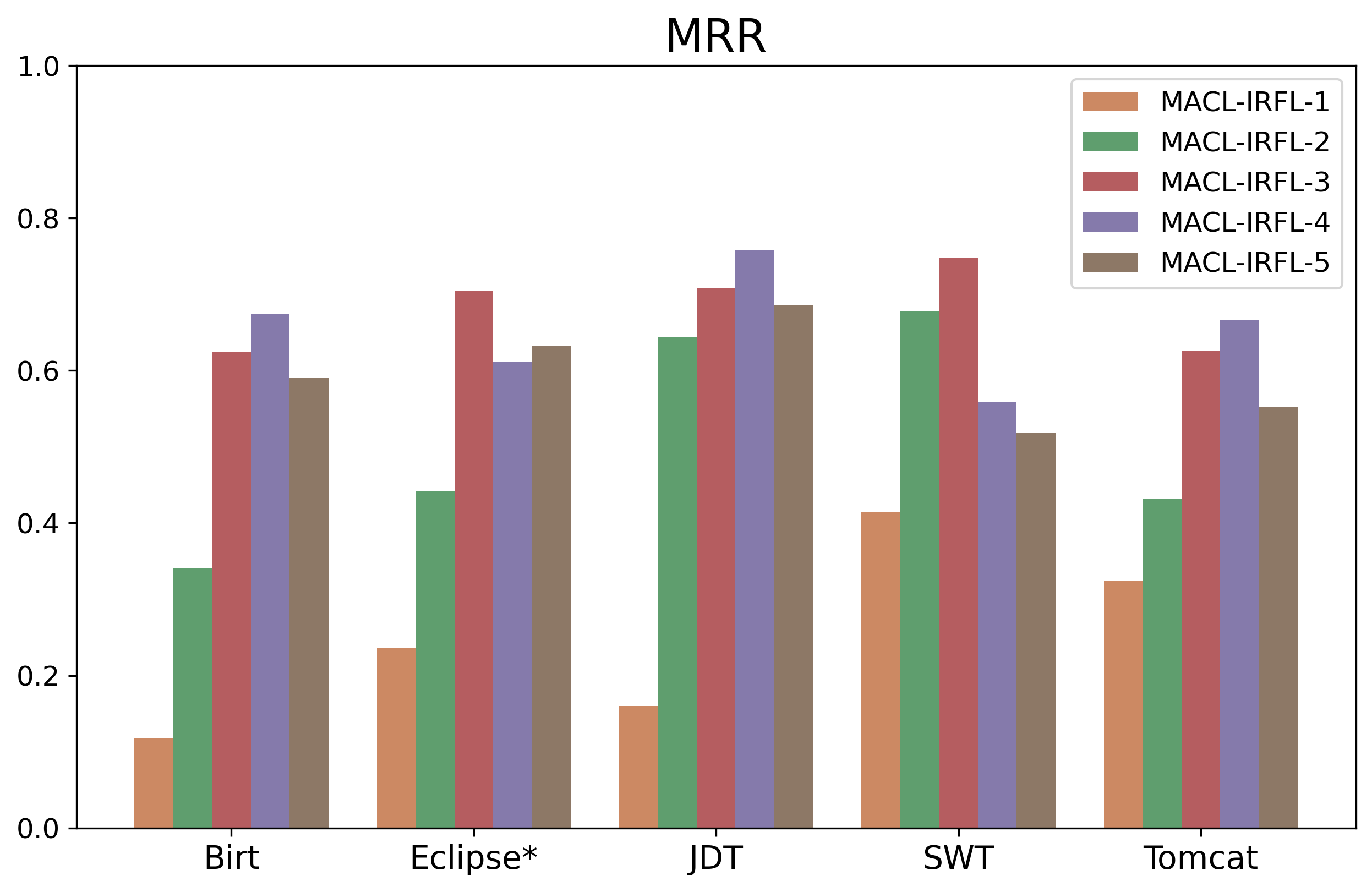}
\end{minipage}%

\caption{Detailed performance of five projects on different aggregation layer numbers.}
\label{RQ3-detail}
\end{figure*}

To investigate whether the high-layer embedding propagation can help improve the performance of MACL-IRFL, we vary the number of layers $L$ of MACL-IRFL from 1 to 5 and compare their performance under different layers. 
Table \ref{RQ3-average} shows the overall performance of the five projects, and MACL-IRFL-$L$ denotes our model with $L$ layers embedding propagation. Fig. \ref{RQ3-detail} illustrates the detailed performance of the five projects at different layer settings.
We have the following observations.

First, \textbf{when the layers $L$ increase from 1 to 3, the performance of MACL-IRFL can be improved substantially.} In particular, MACL-IRFL-2 and MACL-IRFL-3 achieve consistent an improvement over MACL-IRFL-1 on all datasets. For example, MACL-IRFL-2 outperforms MACL-IRFL-1 by 23.81\%, 28.37\%, 30.33\%, 22.89\% and 25.71\% on Accuracy@1, Accuracy@5, Accuracy@10, MAP, and MRR on the five datasets, respectively. MACL-IRFL-3 outperforms MACL-IRFL-1 by 39.94\%, 48.87\%, 48.99\%, 39.32\% and 43.16\% on Accuracy@1, Accuracy@5, Accuracy@10, MAP, and MRR on the five datasets, respectively. 
Moreover, the statistical test results also show that both MACL-IRFL-2 and MACL-IRFL-3 perform significantly better than MACL-IRFL-1 ($p$-value = 0.0079 \textless 0.05, and $|\delta|$ = 1.0 $\geq$ 0.474). \textbf{These improvements mainly benefit from the modeling of the high-order relationships among bug reports and source code files, carried by the high-layer GNN embedding propagation.} This validates the conclusion of previous studies \cite{hamilton2017inductive} that nodes incrementally gain more and more information from further reaches of the graph as aggregation layers or search depth deepens.

Second, we found that \textbf{when the depth of MACL-IRFL $L$ exceeds 3, the improvements are not obvious.} For example, when $L$ exceeds 5, the performance of MACL-IRFL decreases gradually. The statistical test results between MACL-IRFL-3 and MACL-IRFL-4 also show that the $p$-value = 0.6905 ($\geq$ 0.05). It demonstrates that when we increase the number layers $L$ from 3 to 4, there is no significant performance improvement. \textbf{The reason may be that an overly great depth will overfit the model and lowers the performance of MACL-IRFL.} 
As we can see from the details shown in Fig. \ref{RQ3-detail}, the performance of \textit{Birt}, \textit{JDT} and \textit{Tomcat} consistently improves from layer 1 to layer 4, reaching optimal performance at layer 4. The performance of projects \textit{Eclipse Platform UI} and \textit{SWT} has been improving from layer 1 to layer 3, with a slight decline at layer 4. Nevertheless, the experimental results can be generalized that stacking 3 or 4 layers of graph structure of our datasets are sufficient to capture the useful information among bug reports and source code files.

\section{Threats to Validity}\label{threats to validity}
In this section, we discuss three kinds of potential threats that may affect the validity of our study: threats to internal validity, threats to external validity, and threats to construct validity.

One of the key threats to internal validity relates to author bias in the experiments. The dataset that we obtain is taken from prior papers \cite{xiao2017improving, xiao2019improving}, and has been evaluated by other fault localization techniques \cite{zhou2012should, ye2016word, lam2017bug, gharibi2018leveraging, xiao2017improving, xiao2019improving, zhang2020exploiting, qi2021dreamloc, liang2022modeling, zhu2022bl, han2023bjxnet}
The data are bug reports taken from bug tracking system from real projects and thus are realistic. Another key threat is the specific parameter choices we used to build our MACL-IRFL model. To mitigate the impact of this threat, all parameters were either studied by us or were reported in other prior reputable papers as recommended or optimal \cite{zhou2012should, qi2021dreamloc, zhu2022bl}. Therefore, we believe there are limited threats to the internal validity of this study.

The threats to external validity are concerned with the generalizability of our results. In our experiments, we conduct the experiments based on five Java open-source projects. Admittedly, the projects that we analyze may not represent all the projects out there. We are not sure whether our model is effective in other projects or other programming languages. In future work, we need further evaluate the effectiveness and efficiency of the scalable data.

The threats to construct validity relate to the suitability of our evaluation metrics. We have used Accuracy@N, MAP, and MRR as evaluation metrics. These metrics were also used by prior fault localization studies \cite{zhou2012should, ye2016word, lam2015combining, lam2017bug, gharibi2018leveraging, xiao2017improving, xiao2019improving, qi2021dreamloc, zhang2020exploiting, liang2022modeling, han2023bjxnet}, thereby mitigating concerns regarding their suitability and suggesting limited threats to construct validity.
Furthermore, MACL-IRFL is based on GNN. One of the  limitations of GNN-based approaches is the scalability \cite{fan2017big}. Specifically, when dealing with large graphs with a high number of nodes and intricate relationships, GNN-based approaches will encounter challenges due to substantial memory requirements and protracted training durations. To mitigate this threat, we plan to explore subsampling strategies to make our model more applicable in future work.

\section{Related work}\label{related work}
In this section, we briefly review the related work, including text-based fault localization and semantic-based fault localization.

\subsection{Text-based fault localization}
Text-based fault localization techniques predominantly rely on the lexical similarity between bug reports and source code, where the bug report texts is used to formulate  a query that is matched to a corpus of elements, i.e., classes or methods. The classic models employed in IRFL include topic-based models, Vector Space Model (VSM), and so on. These methodologies offer robust frameworks for fault localization efficiently through textual analysis.

The topic-based model assumes that both bug reports and source code are generated by some topic, thereby enabling the similarity between bug reports and source code to be calculated based on their topic distributions. Lukins et al. \cite{lukins2008source, lukins2010bug} applied Latent Dirichlet Allocation (LDA) to extract latent topics from source code files and bug reports, subsequently indexing source files based on these topics. Given a new bug report, they constructed a textual query from its description and searched the indexed source files using the VSM. 
Nguyen et al. \cite{nguyen2011topic} introduced BugScout, which used a specialized LDA-based topic model to represent technical aspects as topics in bug reports and source files. BugScout comprises two components: the S-component models source files, while the B-component, an extended LDA model, considers both a bug report's own topic distribution and the topic distributions of its corresponding buggy source files. It predicts the relevant source files for a new bug report by estimating its topic distribution and comparing it with historical source file distributions, prioritizing files with a history of more defects.

The Vector Space Model (VSM) is the most frequently utilized by many fault localization techniques. VSM is based on the bag-of-words model, which regards bug reports and souce code files as document. Each document is expressed as a vector of token weights typically computed as a product of TF (Term frequency) and IDF (Inverse Document Frequency) of each token. Subsequently,  the cosine similarity is used to determine how close the two vectors are. For instance, Zhou et al. \cite{zhou2012should} proposed BugLocator, which utilized rVSM and considered information about similar bugs that have been fixed before. Firstly, they employed rVSM to search and rank relevant source code files. Subsequently, they collected the similar bugs that have been fixed before, and ranked the relevant files by analyzing past similar bugs and their fixes. Ultimately, they combined the ranks from both sources, utilizing a final score to present a prioritized list of source files, thereby facilitating the efficient localization of new bugs.
Saha et al. \cite{saha2013improving} proposed the BLUiR approach, which divides bug reports into two parts: summary and description, and segments source code into four components: class names, method names, variable names, and comments. Based on a pre-trained BM25 model, these segments are represented as feature vectors. The textual similarity scores between the bug report and each component of the source code are then calculated individually and aggregated to form the final textual similarity score. 

Some researchers transformed the fault localization task into a ranking problem based on parameter learning from an optimization function perspective, specifically optimizing the ranking of the relevance between current source code and a given bug report according to the historical bug report resolution information. Ye et al. \cite{ye2014learning}  introduced an adaptive ranking approach using a learning-to-rank technique and leveraged domain knowledge features in software engineering for locating relevant buggy files, such as API descriptions of library components used in the code, the bug-fixing history, and the code change history. They extracted six features (surface lexical similarity, API-enriched lexical similarity, collaborative filtering score, bug-fixing recency, and bug-fixing frequency) and gave scores to source files by the weighted combination of these features.
Gharibi et al. \cite{gharibi2018leveraging} presented a multi-component fault localization approach that leveraged different textual properties of bug reports and source files as well as relations between previously fixed bug reports and a newly received one. Their approach has five components that are the token matching, VSM similarity, stack trace, semantic similarity, and fixed bug reports. Each of the mentioned components gives a score to source files with respect to each bug report. Then these scores are combined to get a final ranking score that is used to rank relevant source files.

\subsection{Semantic-based fault localization}

However, text-based approaches do not work well if the code contains few common terms with a new bug report. Dealing with lexical mismatches between terms of bug reports and source files is an important subject. More recently, software engineering researchers have been interested in the applying deep learning techniques to bridge the lexical gap between natural language text and source code. 

Huo et al. \cite{huo2016learning} proposed NP-CNN, which leveraged Convolutional Neural Networks (CNNs) to learn unified features from natural language and source code in programming language for bug localization problems. Their approach includes intra-language feature extraction layers and cross-language feature fusion layers. The former employed separate CNNs to extract intra-language features form natural language and programming language, while the latter employed a fully connected neural network to fuse middle-level features extracted from bug reports and source code files to generate a unified feature representation. Finally, the unified feature representation is then passed through a fully connected layer for classification.

Lam et al. \cite{lam2015combining, lam2017bug}, proposed DNNLOC, which combines a kind of deep neural network (Auto-Encoder) and the rVSM to improve bug localization performance. Their approach firstly employed rVSM to capture textual similarity features between bug reports and source code. It further learned the associations between identifiers, API method calls and classes, comments, and API descriptions in the code within bug reports and source code. Finally, a feature fusion layer is utilized to learn the weights to combine all features for an accurate ranking score.

Ye et al. \cite{ye2016word} first leveraged word embedding to project text and code as meaning vectors into a shared representation space to bridge the lexical gap between natural language text and source code. In the proposed architecture, word embeddings are first trained on API documents, tutorials, and reference documents, and then aggregated in order to estimate semantic similarities between documents.

Xiao et al. \cite{xiao2017improving} proposed a deep learning model named DeepLocator, which consisted of an enhanced CNN together with a new rTF-IDuF method and the pre-trained word2vec technique to improve the performance of fault localization. The follow-up work DeepLoc of Xiao et al. \cite{xiao2019improving} used two different word-embedding techniques (Sent2Vec and combined word2vec) to convert bug reports and source files into vectors respectively. Xiao et al. \cite{xiao2018bug} further employed CNN with multiple filters and an ensemble of random forests with multi-grained scanning to extract semantic and structural features from the word vectors derived from bug reports and source files. Finally, a cascade forest consisting of two Completely Random Tree Forests (CRTFs) and two random forests is employed to extract deep features from both bug reports and source code files and perform classification.

Qi et al. \cite{qi2021dreamloc} proposed a framework DreamLoc, which utilized a relevance matching model to locate buggy files. 
DreamLoc comprises of three components, the \textit{Wide} component measured the relevance from the perspective of domain-specific features in software engineering, while the \textit{Deep} component is a deep relevance matching model, which combined local matching and global matching and considers the exact matching of keywords, the local details in source files as well as the difference in importance of words. Finally, the \textit{Fusion} component fused the results of \textit{Wide} and \textit{Deep} components and gets the final output that shows how likely the source file is relevant to the bug report.

Huo et al. \cite{huo2020control} proposed CG-CNN, which 
enhanced the unified features for bug localization by exploiting structural and sequential nature from the control flow graph. Specifically, CG-CNN firstly used CNN and DeepWalk model to learn feature representation of each statement and further extract semantics from control flow graph based on multi-instance decomposition. 

Zhang et al. \cite{zhang2020exploiting} proposed KGBugLocator, which also extracted rich information of source code from the perspective of graph. KGBugLocator solved the fault localization problem via code knowledge graph embeddings, code-specific features and bug-specific features. Specifically, KGBugLocator first exploited interrelation information via code knowledge graph. Subsequently, it employed a keywords supervised bi-directional attention mechanism to mine the interaction information between bug reports and source code files. Ultimately, a fault localization component calculated similarity scores based on the integrated representations.

Liang et al. \cite{liang2022modeling} proposed FLIM, which leveraged CodeBERT to capture function-level code semantics and computed bug report-source code similarity through aggregated function interactions. FLIM consists of two stages. In the first stage, they initially built a training dataset called MIX dataset by splitting a source code file into a set of functions. Following this, they utilized the MIX dataset to fine-tune the CodeBERT model to make it adaptive to fault localization task. In the second stage, they employed the fine-tuned CodeBERT model to predict the relevance between the functions of the source code files and two parts (i.e., summary and description) of the bug report. Finally, semantic features extracted from CodeBERT and IR features are fused to a learning-to-rank model which gives the final relevance score.

Han et al. \cite{han2023bjxnet} proposed a multi-modal representation learning framework bjXnet, which integrated the text semantic features of the source code (shallow semantic features) and the graph features of the CPG (deep semantic features) by using the attention mechanism. Specifically, bjXnet employed the GatedGraphConv layer based on the GGNN to build the GatedGCN network for graph feature extraction, and utilized TextCNN as the text feature extractor to extract the text content of bug reports and source code files. Finally, bjXnet embedded bug reports, source code text, and source code CPGs into the same semantic space, comparing the relationship between report representation vectors and code representation vectors by the cosine similarity to determine whether the report and source code are related.

Apart from focusing solely on semantic matching techniques, some researchers have embarked on endeavors to improve the performance of fault localization by exploring other perspectives and methods combined with advanced semantic models. For instances, Zhu et al. \cite{zhu2022bl} proposed BL-GAN, which leveraged the adversarial training strategy and reinforcement learning for fault localization. In specific, the generator in their model generated code file paths for each given bug report, so that it can further construct pairs of bug report and code file. On the other hand, the discriminator attempted to distinguish whether the pair of bug report and code file is from the generator or not. BL-GAN adopted an attention-based Transformer architecture to capture semantic and sequence information to process bug reports, while incorporated a novel multi-layer GCN to process the source code in a graphical view to capture the proprietary structural information in code files, thereby enhancing the performance of fault localization.

\section{Conclusion}\label{conclusion}
In this paper, we proposed a novel multi-view adaptive contrastive learning approach for fault localization. We formally defined the interaction relationships, co-citation relationships and similarity relationships, and explicitly encoded them into the representation learnings of bug reports and source code files. 
Our approach adopts a multi-task learning framework which can supplement the fault localization loss with extra contrastive learning between report-code, report-report, and code-code views. The contrastive learning task can extract the information share by these views, and thus alleviates the task-irrelevant noises from auxiliary information.
To evaluate the benefits of our model, we conducted extensive experiments on five open-source Java projects. The results show that our model can improve over the best baseline up to 28.93\%, 25.57\% and 20.35\% on Accuracy@1, MAP and MRR, respectively. Our future work will extend our model to support more fine-grained localizing tasks and different programming languages.

\bmhead{Acknowledgements}

This work was partially supported by National Key R\&D Plan of China (Grant No.2024YFF0908003), Natural Science Foundation of China (No.62472326), CCF-Zhipu Large Model Innovation Fund (No.CCF-Zhipu202408), and the Major Program(JD) of Hubei Province (2023BAA018).

\bibliography{sn-bibliography}

\end{document}